\documentstyle[11pt,aaspp4,amssym]{article}

\newcommand{\kms}{${\rm km\ s^{-1}}$}

\received{1 October 1997}
\accepted{15 December 1997}
 
\lefthead{Trager et al.}
\righthead{Old Stellar Populations. IV.}
 
\begin{document}

\title{Old Stellar Populations. VI. Absorption-Line Spectra of
Galaxy Nuclei and Globular Clusters\altaffilmark{1}}

\author{S. C. Trager\altaffilmark{2}}
\affil{UCO/Lick Observatory and Board of Studies in Astronomy and
Astrophysics, \\ University of California, Santa Cruz\\Santa Cruz,
CA 95064\\sctrager{\@@}ociw.edu}
\authoraddr{Carnegie Observatories, 813 Santa Barbara St., Pasadena,
CA 91106; sctrager{\@@}ociw.edu}

\author{Guy Worthey\altaffilmark{3,4}}
\affil{Astronomy Department, University of Michigan\\Ann Arbor, MI
48109-1090\\worthey{\@@}astro.lsa.umich.edu}
\authoraddr{Ann Arbor, MI 48109-1090; worthey{\@@}astro.lsa.umich.edu}

\author{S. M. Faber}
\affil{UCO/Lick Observatory and Board of Studies in Astronomy and
Astrophysics,\\University of California, Santa Cruz\\Santa Cruz, CA
95064\\faber{\@@}ucolick.org}
\authoraddr{Santa Cruz, CA 95064; faber{\@@}ucolick.org}

\author{David Burstein}
\affil{Department of Physics and Astronomy, \\Arizona State
University\\Tempe, AZ 58287-1504\\burstein{\@@}samuri.la.asu.edu}
\authoraddr{Tempe, AZ 58287-1504; burstein{\@@}samuri.la.asu.edu}
 
\author{J. Jes\'us Gonz\'alez}
\affil{Instituto de Astronom{\'\i}a---UNAM\\Apdo Postal 70-264,
M\'exico D.F., Mexico\\jesus{\@@}astroscu.unam.mx}
\authoraddr{Apdo Postal 70-264, M\'exico D.F., Mexico;
jesus{\@@}astroscu.unam.mx}

\altaffiltext{1}{Lick Observatory Bulletin \#1375}
\altaffiltext{2}{Present address: Observatories of the Carnegie
Institution of Washington, 813 Santa Barbara Street, Pasadena, CA
91101}
\altaffiltext{3}{Hubble Fellow}
\altaffiltext{4}{Present address: Department of Physics and Astronomy,
St.~Ambrose University, Davenport, IA 52803-2829}

\begin{abstract}
We present absorption-line strengths on the Lick/IDS line-strength
system of 381 galaxies and 38 globular clusters in the 4000--6400 \AA\
region.  All galaxies were observed at Lick Observatory between 1972
and 1984 with the Cassegrain Image Dissector Scanner spectrograph,
making this study one of the largest homogeneous collections of galaxy
spectral line data to date.  We also present a catalogue of nuclear
velocity dispersions used to correct the absorption-line strengths
onto the stellar Lick/IDS system.  Extensive discussion of both random
and systematic errors of the Lick/IDS system is provided.  Indices are
seen to fall into three families: $\alpha$-element-like indices
(including CN, Mg, Na D, and TiO$_2$) that correlate positively with
velocity dispersion; Fe-like indices (including Ca, the G band,
TiO$_1$, and all Fe indices) that correlate only weakly with velocity
dispersion and the $\alpha$ indices; and H$\beta$ which
anti-correlates with both velocity dispersion and the $\alpha$
indices.  C$_2$4668 seems to be intermediate between the $\alpha$ and
Fe groups.  These groupings probably represent different element
abundance families with different nucleosynthesis histories.
\end{abstract}
 

\section{Introduction}

This paper is the sixth in a series describing a two-decades long
effort to comprehend the stellar populations of early-type galaxies.
Previous papers in this series have defined the Lick/IDS
absorption-line index system, presented observations of globular
clusters and stars, and derived absorption-line index fitting
functions (\cite{BFGK84}; \cite{FFBG85}; \cite{BFG86};
\cite{Gorgas93}).  Worthey et al.\ (1994, hereafter Paper V) expanded
the original eleven-index system to 21 indices and presented the
complete library of stellar data.  Other papers utilizing this
database presented preliminary galaxy Mg$_2$ strengths
(\cite{Burstein88}; \cite{Faber89}), galaxy velocity dispersions
(\cite{FJ76}; \cite{Davies87}; \cite{DalleOre91}), comparisons of
morphological disturbances with absorption-line strengths
(\cite{Schweizer90}), and preliminary comparisons of galaxy
absorption-line strengths with models (\cite{WFG92}; \cite{Worthey92},
1994; \cite{FTGW95}; \cite{WTF96}; Trager 1997).

The Lick/IDS system has also been used extensively by other authors.
Galaxy and globular cluster line strengths on this system have been
published by, among others, Efstathiou \& Gorgas (1985); Couture \&
Hardy (1988); Thomsen \& Baum (1989); Gorgas, Efstathiou \& Aragon
Salamanca (1990); Bender \& Surma (1992); Davidge (1992); Guzm{\'a}n
et al.\ (1992); Gonz{\'a}lez (1993); Davies, Sadler \& Peletier
(1993); Carollo, Danziger \& Buson (1993); de Souza, Barbuy \& dos
Anjos (1993); Gregg (1994); Cardiel, Gorgas \& Aragon-Salamanca
(1995); Fisher, Franx \& Illingworth (1995, 1996); Bender, Zeigler \&
Bruzual (1996); Gorgas et al.\ (1997); J{\o}rgensen (1997); Vazdekis
et al.\ (1997); Kuntschner \& Davies (1997); and Mehlert et al.\
(1997).  Much theoretical and empirical calibration of the Lick/IDS
absorption-line strengths of stars (particularly Mg$_2$) has also been
pursued by, e.g., Gulati, Malagnini \& Morossi (1991, 1993); Barbuy,
Erdelyi-Mendes \& Milone (1992); Barbuy (1994); McQuitty et al.\
(1994); Borges et al.\ (1995); Chavez, Malagnini \& Morossi (1995);
Tripicco \& Bell (1995); and Casuso et al.\ (1996).  The Lick/IDS
indices of the stellar populations of composite systems have been
modelled by, e.g., Aragon, Gorgas \& Rego (1987); Couture \& Hardy
(1990); Buzzoni, Gariboldi, \& Mantegazza (1992); Buzzoni, Mantegazza
\& Gariboldi (1994); Matteucci (1994); Buzzoni (1995); Weiss, Peletier
\& Matteucci (1995); Tantalo et al.\ (1996); Bressan, Chiosi \&
Tantalo (1996); Bruzual \& Charlot (1996); de Freitas Pacheco (1996);
Vazdekis et al.\ (1996); Greggio (1997); and M{\"o}ller, Fritze-von
Alvenleben \& Fricke (1997).  We also point out the ongoing efforts of
Rose and colleagues to study old stellar populations using
high-resolution absorption-line strengths in the blue (Rose 1985a,
1985b, 1985c, 1994; Rose \& Tripicco 1986; Rose, Stetson \& Tripicco
1987; Bower et al.\ 1990; Caldwell et al.\ 1993, 1996; Rose et al.\
1994; Leonardi \& Rose 1996; Caldwell \& Rose 1997), and those of
Brodie, Huchra and colleagues to study extragalactic globular cluster
systems using a spectrophotometric index system in the red (Brodie \&
Huchra 1990, 1991; Huchra, Kent \& Brodie 1991; Perelmuter, Brodie \&
Huchra 1995; Huchra et al.\ 1996).

The full IDS database contains absorption-line strengths of 381
galaxies, 38 globular clusters, and 460 stars based on 7417 spectra
observed in the 4000--6400 \AA\ region.  Here, we present final IDS
index strengths for galaxies and globular clusters.  All were observed
at Lick Observatory between 1972 and 1984 with the Cassegrain Image
Dissector Scanner spectrograph, making this study one of the largest
homogeneous collections of galaxy spectral line data to date.

This paper begins by describing the method of measuring Lick/IDS
absorption-line strengths in Section~\ref{sec:2.2}.
Section~\ref{sec:2.3} presents a discussion of uncertainties in these
measurements.  As early-type galaxies typically have significant
internal motions, Section~\ref{sec:2.4} derives the corrections needed
to bring the galaxies to a common zero-ve\-lo\-ci\-ty-dis\-per\-sion
system and the additional uncertainties incurred by this correction.
Section~\ref{sec:2.4} also presents the velocity dispersions
themselves and a first discussion of ``families'' of absorption-line
indices according to each index's behavior with velocity dispersion.
Section~\ref{sec:syst} illustrates remaining levels of suspected
systematic errors and compares to previously published values.
Finally, Section~\ref{sec:2.5} presents final mean corrected indices
and their associated errors for the entire sample.

This paper plus Paper V (for the stellar data) together contain the
sum total of all observations on the Lick/IDS system.  Previously
published data on galaxies and globular clusters are superseded by the
values given here.  Table~\ref{tbl1} presents published papers
containing IDS data, the index data presented in those papers, the
index measurement method, and the run corrections applied (terms are
explained in Sections~\ref{sec:2.2} and \ref{sec:syst}).

\placetable{tbl1}

The complete tables in this paper and individual spectra for all
Lick/IDS stellar, globular cluster, and galaxy observations are
available electronically from the Astrophysical Data Center
(http://adc.gsfc.nasa.gov/adc.html).  The complete versions of the
long tables (Tables~\ref{tbl4}, \ref{tbl6a}--\ref{tbl6d}) are also
presented in the electronic Astrophysical Journal Supplements.

\section{Absorption-Line Measurements}
\label{sec:2.2}

A general introduction to the Image Dissector Scanner (IDS) is given
by Robinson \& Wampler (1972), and further relevant details are found
in Faber \& Jackson (1976), Burstein et al.\ (1984) and Faber et al.\
(1985).  A discussion of signal-to-noise and the noise power spectrum
is presented in Faber \& Jackson (1976) and Dalle Ore et al.\ (1991).

\begin{figure}
\plotone{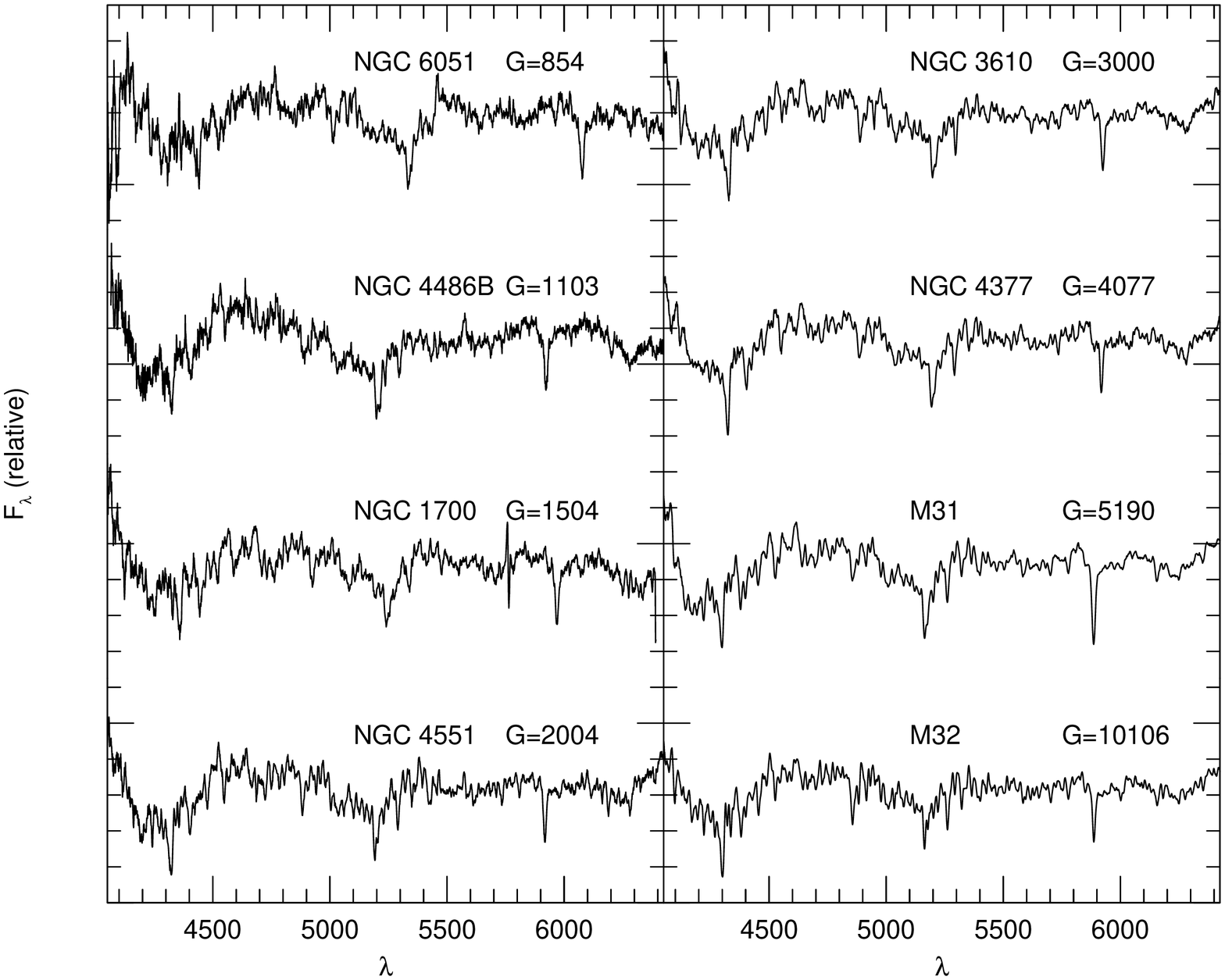}
\caption{A selection of IDS spectra covering a range of
S/N. Spectra are labelled with their name and goodness $G$ (see
Section 3).\label{fig1}}
\end{figure}

Briefly, spectra were obtained between 1972 and 1984 using the
red-sensitive IDS and Cassegrain spectrograph on the 3m Shane
Telescope at Lick Observatory.  The spectra cover roughly
4000--6400~\AA\ and have a resolution of about 9~\AA\ (about 30\%
higher at the ends of the region) although this varied slightly from
run to run.  Most spectra of galaxy nuclei were taken through a
spectrograph entrance aperture of $1\farcs 4$ by $4\arcsec$, with a
second aperture for sky subtraction located $21\arcsec$ or $35\arcsec$
away.  Object and sky were chopped between these apertures in such a
way as to equalize the time spent in each.  Long slit observations of
galaxies (of width $1\farcs 4$ and various lengths) and spatial scans
of globular clusters and galaxies were also taken and have equivalent
resolution to the nuclear data.  Larger-aperture observations of
galaxies with wider slits (typically off-nucleus observations or dwarf
galaxies) were also taken and calibrated separately.  These wide-slit
observations have lower spectral resolution.  Helium, neon, mercury,
and (in later observations) cadmium lamps provided wavelength
calibrations at the beginning and end of every night.  Global shifts
and stretches of the wavelength scale of up to 3~\AA\ per observation
could occur due to instrument flexure and variable stray magnetic
fields.  Spectra were not fluxed, but rather divided by a
quartz-iodide tungsten lamp, the energy distribution of which was made
more constant with wavelength by ``rocking'' the dispersion grating in
a reproducible, systematic manner.  Line-strength standard stars
(detailed in Paper V) were observed nightly to insure a calibration of
the system.  A sampling of low- to high-quality galaxy nuclei
spectra are shown in Figure~\ref{fig1}.  For display purposes, these
spectra have been flattened by a fifth-order polynomial fit.  However,
all measurements of line strengths were made on the original,
unflattened spectra.

\placefigure{fig1}

\subsection{The Lick/IDS System}

The Lick/IDS absorption-line index system is fully described in Paper
V.  We present a summary here and point out changes to the system
caused by measuring galaxies with significant systemic velocities and
velocity dispersions.  Absorption-line strengths are measured in the
Lick/IDS system by ``indices,'' where a central ``feature'' bandpass
is flanked to the blue and red by ``pseudocontinuum'' bandpasses.  The
choice of bandpasses is dictated by three needs: proximity to the
feature, less absorption in the continuum regions than in the central
bandpass, and maximum insensitivity to velocity dispersion broadening.
While the last point is unnecessary when measuring stars, in the case
of galaxies it is crucial, and it sets a minimum length for the
pseudocontinuum bandpasses.  The sidebands are called
``pseudocontinua'' because the resolution of the Lick/IDS system does
not allow the measurement of ``true'' continua in late-type stars or
in most galaxies.

Table~\ref{tbl2} presents the bandpasses of the 21 Lick/IDS absorption
line indices and the features measured by these indices.  The
wavelengths have been further refined since Paper V through
cross-correlation with more accurate CCD spectra taken by GW.  Indices
1--8 have been corrected by $1.25$~\AA, and indices 17--21 have been
corrected by $1.75$~\AA.  Uncertainties of 0.3~\AA\ are still present
in these bandpass definitions, but such shifts produce negligible
changes in the measured indices.  Systemic velocities of the galaxies
sometimes caused the reddest absorption features to fall outside of
the wavelength range of the observation, and the starting wavelength
of the spectra also varied somewhat.  Occasionally other effects
prevented the measurement of particular indices, including bubbles in
the immersion oil of the IDS, exceptionally strong galaxy emission,
or poorly-subtracted night sky lines (see Section~\ref{sec:2.5} for a
complete list).  As a result, not all indices are measured for all
galaxies.

\placetable{tbl2}

The Lick/IDS index system was nominally designed to include six
different molecular bands [CN4150, the G band (CH), MgH, MgH $+$ Mg
$b$, and two TiO bands] plus 14 different blends of atomic absorption
lines.  The CN$_2$ index, introduced in Paper V, is a variant of the
original CN$_1$ index with a shorter blue sideband to avoid H$\delta$.
Along with the higher-order Balmer lines presented in Worthey \&
Ottaviani (1997), we believe we have extracted all of the useful
absorption features from the Lick/IDS stellar and galaxy spectra.

We note here the recent work of Tripicco \& Bell (1995), who modelled
the Lick/IDS system using synthetic stellar spectra.  They found that
many of the Lick/IDS indices do not in fact measure the abundances of
the elements for which they were named.  Column 6 of Table~\ref{tbl2}
describes their results, in order of the most significant contributing
element.  To retain conformity with previously published studies we
have chosen not to rename most of the Lick/IDS indices for their
primary contributor.  However, following Worthey, Trager \& Faber
(1996), we have renamed the Fe4668 index {\it C$_{\rm 2}$4668}.

\subsection{Index measurements}
\label{sec:2.2b}

Index measurements from the Lick/IDS galaxy spectra are problematic
owing to unpredictable wavelength shifts and stretches (of order
1--3~\AA) and also from the (sometimes unknown) systemic radial
velocities of the galaxies themselves.  Indices were measured
automatically using the program AUTOINDEX written by
J. J. Gonz{\'a}lez and G. Worthey.  This program begins by locating Na
D (centroid assumed at 5894~\AA) and the G band (centroid assumed at
4306~\AA) or, for a few galaxies with very strong Balmer lines,
H$\gamma$ (centroid assumed at 4340~\AA). It then removes any global
wavelength shift and stretch, including the effects of the systemic
velocity.  Local wavelength shifts at each index are calculated by
cross-correlating the galaxy spectrum with a template spectrum in
the region around each index.  For galaxies, a K0 giant template is
generally used, but occasionally an F5 dwarf template is used for
galaxies with very strong Balmer lines.

Once each index was centered, it is measured following the scheme
outlined in Paper V.  The mean height in each of the two
pseudocontinuum regions is determined on either side of the feature
bandpass, and a straight line was drawn through the midpoint of each
one.  The difference in flux between this line and the observed
spectrum within the feature bandpass determines the index.  For narrow
features, the indices are expressed in {\AA}ngstroms of equivalent
width; for broad molecular bands, in magnitudes.  Specifically, the
average pseudocontinuum flux level is
\begin{equation}
F_P=\int_{\lambda_1}^{\lambda_2}F_\lambda d\lambda/(\lambda_2 -
\lambda_1),
\end{equation}
where $\lambda_1$ and $\lambda_2$ are the wavelength limits of the
pseudocontinuum sideband.
If $F_{C\lambda}$ represents the straight line connecting the
midpoints of the blue and red pseudocontinuum levels, an equivalent
width is then
\begin{equation}
{\rm EW}=\int_{\lambda_1}^{\lambda_2}\left(1-{F_{I\lambda}\over
F_{C\lambda}}\right) d\lambda, \label{eq:ew}
\end{equation}
where $F_{I\lambda}$ is the observed flux per unit wavelength and
$\lambda_1$ and $\lambda_2$ are the wavelength limits of the feature
passband.  Similarly, an index measured in magnitudes is
\begin{equation}
{\rm Mag} = -2.5 \log \left[\left({1\over \lambda_2 -
\lambda_1}\right)\int_{\lambda_1}^{\lambda_2}{F_{I\lambda}\over F_{C\lambda}}
d\lambda\right]. \label{eq:mag}
\end{equation}

As explained in Paper V, the above AUTOINDEX definitions differ
slightly from those used in Burstein et al.\ (1984) and Faber et al.\
(1985) for the original 11 IDS indices.  In the original scheme, the
continuum was taken to be a horizontal line over the feature bandpass,
at the level $F_{C\lambda}$ taken at the midpoint of the bandpass.
This flat rather than sloping continuum induces small, systematic
shifts in the feature strengths, as described in further detail in
Section~\ref{sec:syst}.  For now it is sufficient to note that slight
additive corrections have been applied to the new indices to preserve
agreement with the older published data.  These corrections are
discussed in Section~\ref{sec:syst} and are always quite small.

Run corrections for the galaxies also differ from those described for
stars in Paper V.  Stars always have nearly zero velocities, and their
features occupy the same IDS channels on a given run.  It was
therefore found to be advantageous to apply small additive corrections
to all indices to correct for small variations in continuum shape
and/or resolution for that run.  Galaxies however occupy different
channels due their varying radial velocities, making the
stellar-derived continuum shape corrections invalid.  Hence, the
following scheme was adopted, according to the velocity offset of a
galaxy from the stars: globular clusters and galaxies with $cz \leq
300$ \kms\ (i.e., Local Group galaxies) had stellar run corrections
applied for all indices; galaxies with $300 < cz < 10\,000$ \kms\ had
stellar run corrections applied only to the broad molecular indices
measured in magnitudes; and galaxies with $cz \geq 10\,000$ \kms\ had
no run corrections applied.

\section{Error Estimation}
\label{sec:2.3}

The errors of the IDS indices are due partly to photon statistics and
partly to the fact that the flat-field calibration of the IDS had
limited accuracy.  A thorough knowledge of the errors is essential to
the proper use of these data.  The error estimates derived here will
be used in later papers to simulate the absorption-line data and test
the significance of any conclusions.

The IDS was not a true photon-counting detector.  This makes
estimation of uncertainties difficult, as the errors are not strictly
photon counting statistics.  We present in this section a brief
overview of the steps required to derive reasonable error estimates
for galaxy Lick/IDS index measurements.  A complete discussion of
the error estimates presented here may be found in Trager (1997).

In the IDS, light from the spectrograph fell on a series of three
image-tube photocathodes, which amplified the signal by about $10^5$.
The amplified light fell on a phosphor screen, which held the light
long enough for an image dissector to scan and digitize the image
before it faded (\cite{RW72}).  Each incident photon produced a
burst of typically seven to ten detected photons covering $\sim
9$~\AA\ (7 channels) in the digitized scan.  Uncertainties in the
spectra arise from three sources: (1) input photon shot noise, (2) the
statistics of the amplification process, and (3) flat-fielding errors.
This last noise source is due to the movement of the spectrum of the
first photocathode, caused by instrument flexure, and movement of the
amplified spectrum, caused by stray magnetic fields affecting the
magnetically-focused image-tubes and image dissector.  As a result,
flat-field spectra taken at different telescope locations and position
angles do not divide perfectly but rather show low-level undulations a
few channels wide.

The effect of these three noise sources on the power spectrum is
discussed in Dalle Ore et al.\ (1991), Paper V, and Trager (1997).  At
low frequencies the noise is dominated by flat-fielding errors (at
high counts) and photon shot-noise and the statistics of the IDS burst
amplification process (at low counts).  At high frequencies the noise
is dominated by flat-fielding errors (very high counts) and photon
statistics (low and moderate counts).  The resultant power spectrum
changes shape with count level, as shown schematically by Trager
(1997).  In galaxy spectra, photon statistics tend to be the overall
dominant noise source, as opposed to the stellar spectra (Paper V) and
the highest-signal-to-noise galaxies ({\it e.g.,} M31 and M32), in
which flat-fielding errors dominate.

The net result is that the high-frequency noise is a good measure of
photon statistics except at very high count levels, where
flat-fielding errors begin to dominate.  Paper V therefore defined a
``goodness parameter'' that measures the noise power at high
frequencies.  For each spectrum, a Fourier transform was taken of the
256 channels starting at 5519~\AA\ in the rest frame, a region
relatively free of spectral lines.  The average power at high spatial
frequencies was measured, then divided by the power at zero frequency.
The square root of this ratio is a measure of photon noise, and its
inverse is defined to be the goodness $G$.

$G$ is defined such that, if all noise were photon statistics, $G$
would be exactly proportional to $\sigma^{-1}$.  The constant of
proportionality is unknown {\it a priori} (it depends on the average
number of detected photons per burst, which is not well known) but can
be determined empirically by comparing to errors derived from
multiply-observed data.  At high count levels, $G$ saturates (bottoms
out) due to the influence of flat-field errors, and the relation of
$G$ to $\sigma^{-1}$ becomes non-linear.  This curvature can also be
determined empirically from multiply-observed objects.

The empirical calibration proceeds as follows.  Because $G$ scales as
$\sigma^{-1}$ for poor data, it should average quadratically for
multiple observations, and thus we compute the goodness $\langle G
\rangle_k$ of a {\it single,} typical observation of galaxy $k$ as
\begin{equation}
\langle G \rangle_k^{2}={1\over N}\sum G_{i,k}^{2}, \label{eq:meang}
\end{equation}
where $G_{i,k}$ is the goodness of each individual spectrum and $N$ is
the number of observations of galaxy $k$.  $\langle G \rangle_k$ would
be the goodness of each {\it single\/} observation of galaxy $k$ if
all observations were of equal quality.  All galaxies with three or
more observations had average goodnesses computed by
Equation~\ref{eq:meang}.  The same galaxies also had mean standard
deviations computed for each of the 16 Lick/IDS indices between the G
band and Na D (in the spectral range of virtually all galaxies).  The
average total error $\sigma_{{\rm TOT},k}$ of galaxy $k$ averaged over
these 16 indices is calculated as
\begin{equation}
\sigma^2_{{\rm TOT},k} = {1\over 16} \sum^{19}_{j=4}
\left({\sigma_{jk}\over\sigma_{sj}}\right)^2,
\end{equation}
where $j$ is the IDS index number (Table~\ref{tbl2}), $\sigma_{jk}$ is
the standard deviation per observation of index $j$ for galaxy $k$,
and $\sigma_{sj}$ is the standard star error of index $j$
(Table~\ref{tbl2}).  Thus $\sigma_{{\rm TOT},k}$ is the average error
in units of the standard star error for a typical {\it single}
observation of galaxy $k$.  It is an external error determined from
multiple, independent observations of the same object.

\begin{figure}
\plotone{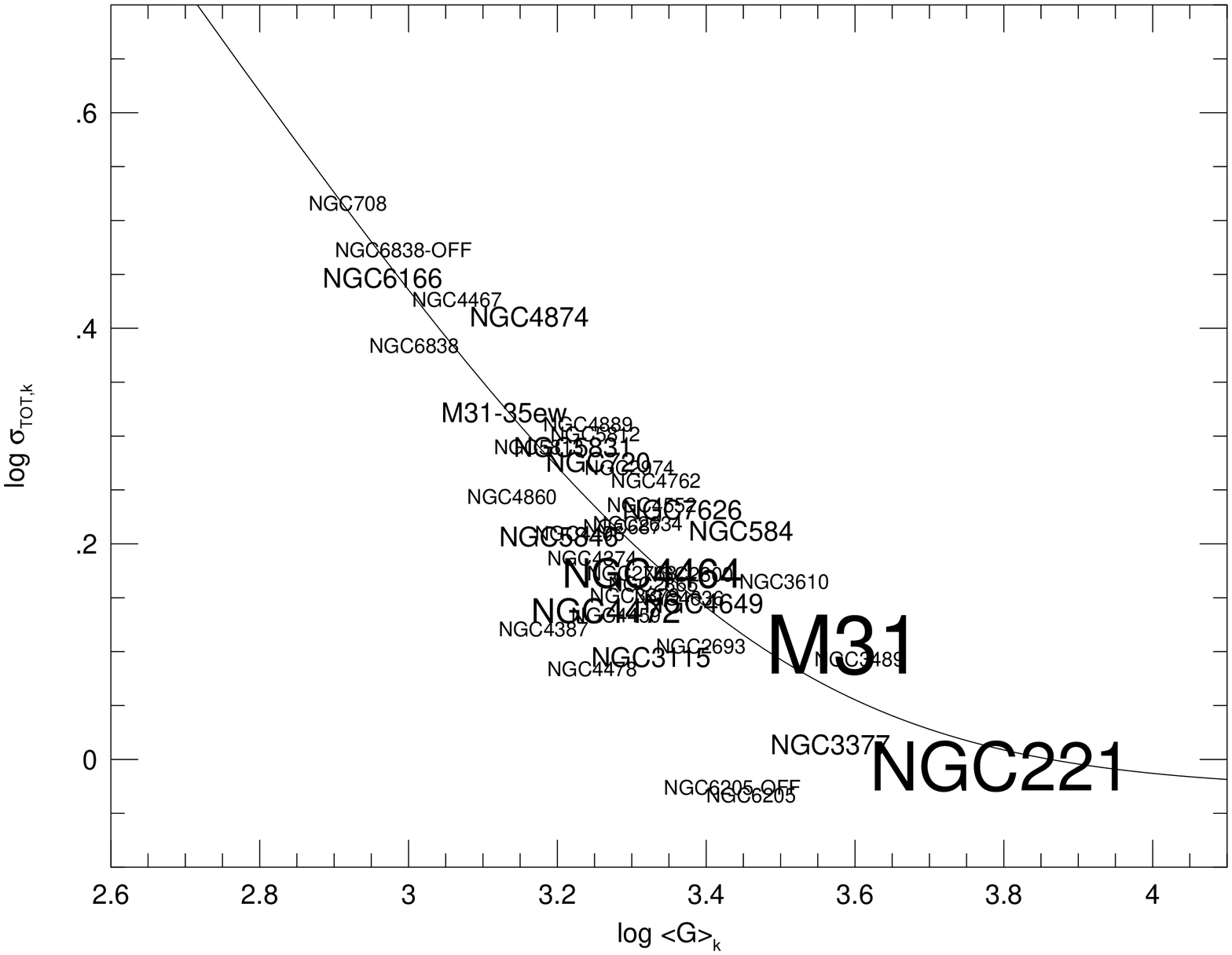}
\caption{Preliminary calibration of the independently determined
error $\sigma_{{\rm TOT},k}$ with goodness $\langle G\rangle_k$.  The
size of the galaxy labels is proportional to the number of
observations.  The middle of the label is the location of the point.
The relation flattens at high $\langle G\rangle_k$ due to
flat-fielding errors.  The solid line is a least-squares linear fit to
the relation $\sigma^2_{{\rm TOT},k} = a \langle G \rangle_k^{-2} + b$
(see Section 3).\label{fig4}}
\end{figure}

To determine a preliminary scaling of total error with goodness,
individual total errors $\sigma_{{\rm TOT},k}$ were plotted against
average goodnesses $\langle G \rangle_k$ (Figure~\ref{fig4}).  There
is a reasonably tight relation between the two with the expected
trend.  The slope is $-1$ in the low-signal limit, where photon
statistics dominate, and flattens out in the high-signal limit, where
flat-fielding errors dominate.  The solid curve is a least squares fit
to the equation $\sigma^2_{{\rm TOT},k} = a \langle G \rangle_k^{-2} +
b$:
\begin{equation}
\sigma^2_{{\rm TOT},k} = \left({2561\over \langle G
\rangle_k}\right)^2 + (0.94)^2. \label{eq:gsig}
\end{equation}

\placefigure{fig4}

Equation~\ref{eq:gsig} is assumed to hold for each individual
observation, with $\langle G \rangle_k$ replaced by $G_{i,k}$.
Weighted mean indices for multiply-observed galaxy $k$ are calculated
as
\begin{equation}
\langle I_{jk}\rangle = \left(\sum_{i=1}^N I_{ijk}/\sigma^2_{{\rm
TOT},i,k} \Bigl/ \sum_{i=1}^N 1/\sigma^2_{{\rm TOT},i,k}\right),
\label{eq:meani}
\end{equation}
where $i$ represents an individual observation, $j$ represents a given
index, and $\sigma_{{\rm TOT},i,k}$ is now derived from $G_{i,k}$
using Equation~\ref{eq:gsig}.

Finally, the error of each mean index $j$ is taken to be
\begin{equation}
\sigma_{jk} = \frac{1}{\sqrt{N}}\sigma_{{\rm TOT},k} \times
\sigma_{sj}, \label{eq:err1}
\end{equation}
where $N$ is the number of observations of galaxy $k$, $\sigma_{sj}$
is the standard star error, and the mean error of galaxy $k$ for all
indices, $\sigma_{{\rm TOT},k}$, is calculated from
Equation~\ref{eq:gsig}, using $\langle G \rangle_k$ determined as in
Equation~\ref{eq:meang}.  This version of $\sigma_{jk}$ is more
accurate than the individual index standard deviations because it uses
the {\it average\/} error per spectrum, $\sigma_{{\rm TOT},k}$, made
possible by knowing the {\it ratios\/} of the errors between indices
from the standard stars.

We then set out to check the quality of these preliminary error
estimates.  We were interested in both the magnitude of the errors
averaged over all indices and the ratio of the individual index
errors.  To anticipate the results, we found that the mean magnitude
of the galaxy errors was well determined (to within 8\%) but that
certain individual error ratios needed adjustment.

The details of this step are given in Trager (1997) but a brief
description follows.  Independent nuclear data from galaxies in the
sample of Gonz{\'a}lez (1993; hereafter G93) were compared against
individual observations of 37 IDS galaxies in common.  The G93 spectra
cover only the region 4780--5600 \AA, and so only the indices from
H$\beta$ through Fe5406 could be compared.  A chi-squared analysis was
performed to determine the relative scaling of the Lick/IDS galaxy
errors with respect to G93.  Gonz{\'a}lez's indices are so accurate
(except for Mg$_1$ and Mg$_2$) that his errors contribute negligibly,
and the resultant $\chi^2$ values are a good test of the Lick/IDS
errors alone.  Though we expect the errors in G93 to be negligible, we
allowed for mean zeropoint and slope differences, as Gonz{\'a}lez
could not calibrate his CCD system precisely onto the IDS system (see
his Figure 4.4).  The error rescalings determined from the G93
comparison were fairly small, about $0.92$.  The exception was Fe5270,
which required a large error rescaling (0.75; {\it i.e.,} the
preliminary IDS errors from Equation~\ref{eq:err1} above were too
large by 25\% in this index).

A further check for wavelengths not covered by the G93 spectra was
performed using pairs of indices from the Lick/IDS sample itself.
Indices were chosen that might be expected {\it a priori} to track
each other closely ({\it i.e.,} to be multiples of one another) and
have similar velocity dispersion corrections.  The best choices came
from the Fe-peak family of indices (see Section~\ref{sec:families},
although note that these indices do not all track Fe abundance---see
Tripicco \& Bell 1995 and Table~\ref{tbl2}).  Two groups were defined
by their similar velocity dispersion corrections: Fe4383, Fe4531, and
Fe5709 were compared against Fe5270; and Fe5782 and Ca4455 were
compared against Fe5335.  The errors of Fe5270 and Fe5335 were first
rescaled to match G93 as described above.  Chi-squared analyses were
performed, and the resultant reduced-$\chi^2$ value was forced to
equal unity by rescaling the Lick/IDS errors of the dependent index.
The error rescalings from these internal comparisons are comparable to
those derived from the G93 comparison, typically again about $0.92$.

A final mean fractional error scaling was then computed from all
scalings derived in these tests.  This mean scaling was again $0.92$.
Errors in the remaining 10 indices were rescaled by this factor.  We
checked the final adopted index scalings by performing a final set of
chi-squared tests on various Fe-line pairs.  The resulting
reduced-$\chi^2$ values were consistent with our final scaling of the
errors to typically within a few percent (and never worse than 5\%).
From these various tests, we believe that systematic errors in the
final uncertainties are $\la 5\%$.

The adjusted final errors for the raw indices of all galaxies and
globular clusters are computed as
\begin{equation}
\sigma_j^{adj} = c_j\sigma_j, \label{eq:gerr}
\end{equation}
where $\sigma_j$ is the preliminary error of index $j$ computed in
Equation~\ref{eq:err1}, and $c_j$ is the scaling of index $j$ relative
to the standard star indices as determined in these tests.  Adopted
values of $c_j$ are shown in Table~\ref{tbl3}.

\placetable{tbl3}

\section{Velocity Dispersion Corrections}
\label{sec:2.4}

The observed spectrum of a galaxy is the convolution of the integrated
spectrum of its stellar population by the instrumental broadening and
the distribution of line-of-sight velocities of the stars.  The
instrumental and velocity-dispersion broadenings broaden the spectral
features, causing the absorption-line indices to appear weaker than
they intrinsically are.  In this section, we discuss the corrections
required to remove the effects of velocity dispersion from the
galaxy index measurements and the additional uncertainties that
arise from these corrections.

\subsection{Velocity dispersion data}

The adopted nuclear galaxy velocity dispersions, their fractional
errors, and their sources are presented in Table~\ref{tbl4}.  The
majority of the velocity dispersions was derived directly from the IDS
spectra themselves.  The basic method was discussed in Dalle Ore et
al.\ (1991), and the data were presented in Davies et al.\ (1987, as
tabulated by Faber et al.\ 1989) and Dalle Ore et al.  Other sources
of nuclear dispersions include G93, the compilation of Faber et al.\
(1997), and the compilation of Whitmore, McElroy \& Tonry (1985), in
order of preference.  The velocity dispersions of both Whitmore et
al.\ and Faber et al.\ are derived from comprehensive literature
searches, but the data of G93 are excellent and uniform (and supersede
all other measurements when available).  Two other sources noted in
Table~\ref{tbl4} (Bender, Paquet \& Nieto 1991; Peterson \& Caldwell
1993) were used for dwarf galaxies.  For a few galaxies, no velocity
dispersions were available, so educated guesses were made by eye or by
comparing against similar galaxies with known velocity dispersions
These rough velocity dispersions are derived for the purpose of
velocity dispersion corrections only and should not be used for any
other purpose.  They are indicated in Table~\ref{tbl4} as Source 8.

\placetable{tbl4}

For off-nuclear observations of galaxies (Table~\ref{tbl6d}), velocity
dispersions were calculated as
\begin{equation}
\sigma_r = \sigma_0\left({r\over1\farcs4}\right)^{-0.06},
\end{equation}
where $r$ is the radius at which the aperture was placed and
$\sigma_0$ is the velocity dispersion given in Table~\ref{tbl4}.  The
exponent is a mean for early-type galaxies as determined from Figure
6.10 of G93.
  
Finally, a few galaxy nuclei were observed by scanning a long slit of
dimensions $1\farcs4 \times 16\arcsec$ across the nucleus to create a
$16\arcsec \times 16\arcsec$ aperture (denoted ``scan'' in
Table~\ref{tbl6d}).  These were observed to determine aperture
corrections to velocity dispersion and Mg$_2$ in Davies et al.~(1987).
For these we have used the velocity dispersions as corrected by
Equation 1 of Davies et al.

\subsection{Corrections from broadened stellar spectra}
\label{sec:2.4b}

To correct absorption-line strengths for the effects of velocity
dispersion, a reference velocity dispersion must be chosen.  As we
plan to compare the indices derived in this study to
stellar-population models based on our stellar observations (Paper V,
Worthey 1994), the indices are corrected to zero velocity
dispersion.  To achieve this goal, a variety of stellar spectra was
convolved with broadening functions of various widths.  A selection of
G dwarfs and the K giant standard stars was convolved with Gaussians
of widths ranging up to $\sigma=450$ \kms.  Index
strengths were measured from each convolved spectrum and compared to
the original strengths.  A third-order polynomial was then fit to the
ratios (ori\-ginal/con\-vol\-ved) for all the stars in each index {\it
versus\/} velocity dispersion.  Several observations of M32 were also
included in the fits (M32 has a very small velocity dispersion
compared to the resolution of the IDS system).  Figure~\ref{fig7}
shows the results of these fits, and Table~\ref{tbl5} presents the
coefficients of the polynomials.

\placefigure{fig7}
\placetable{tbl5}

A velocity-dispersion corrected index is then
\begin{equation}
I_{j,k}^{corr} = C_j(\sigma_v)\times \langle I_j \rangle_k,
\label{eq:icor}
\end{equation}
where $\langle I_{j} \rangle_k$ is the mean value of index $j$ of
galaxy $k$ from Equation~\ref{eq:meani}, and $C_j(\sigma_v)$ is the
velocity-dispersion correction:
\begin{equation}
C_j(\sigma_v) = \sum_{i=0}^3 c_{ij} \sigma_v^i, \label{eq:vcor}
\end{equation}
where $c_{ij}$ are the coefficients of the correction polynomial for
index $j$ (Table~\ref{tbl5}), and $\sigma_v$ is the velocity
dispersion.

Figure~\ref{fig7} shows that considerable scatter exists in certain
velocity-dispersion corrections.  As noted by G93, a variation with
spectral type is seen in several indices.  Some of the scatter is
negligible, reflecting variations in indices that are intrinsically
small (Mg$_1$, TiO$_1$, TiO$_2$).  Scatter in CN$_1$, CN$_2$, and
H$\beta$ is real.  However, CN is not heavily used, while the scatter
in H$\beta$ is inflated due to the inclusion of a few very cool K
giants with H$\beta$ strengths weaker than typical galaxies.  In what
follows, we do not assign any uncertainty to the velocity dispersion
corrections.  The uncertainty in the H$\beta$ correction will be noted
in future papers when applicable.

\subsection{Final errors}

The velocity-dispersion corrections increase the raw index errors,
$\sigma_j$, by the value of the multiplicative correction.  An
additional source of uncertainty is introduced by errors in the
velocity dispersion estimates themselves.  It proves simplest to
discuss these effects in terms of the fractional error of the final
index.

The uncertainty from the velocity dispersion error is computed as the
fractional uncertainty of the galaxy's velocity dispersion multiplied
by the derivative of the correction function at that velocity
dispersion:
\begin{equation}
\sigma_{v,j} = \epsilon_{\sigma_v} {d\ln C_j\over d\ln\sigma_v},
\label{eq:vcerr}
\end{equation}
where $\sigma_{v,j}$ is the fractional uncertainty in the
velocity-dispersion correction of index $j$, $\epsilon_{\sigma_v}$ is
the fractional uncertainty of the velocity dispersion estimate, $C_j$
is the velocity-dispersion correction of index $j$
(Equation~\ref{eq:vcor}), and $\sigma_v$ is the velocity dispersion.
This uncertainty is added in quadrature with the raw fractional error
in the index $j$,
\begin{equation}
\sigma_{f,j}^2=\sigma_{v,j}^2 + \left({\sigma_j^{adj}\over\langle
I_j\rangle}\right)^2,
\label{eq:err2}
\end{equation}
where $\sigma_{f,j}$ is the final fractional uncertainty of index $j$,
$\sigma_j^{adj}$ is the raw error of index $j$
(Equation~\ref{eq:gerr}), and $\langle I_j\rangle$ is the value of
index $j$ uncorrected for velocity dispersion.  The final fractional
error is then multiplied by the velocity-dispersion corrected index
$j$, $I_j^{corr}$ (Equation~\ref{eq:icor}), to determine the final,
corrected error of index $j$:
\begin{equation}
\sigma_j^{corr} = \sigma_{f,j} \times I_j^{corr}. \label{eq:err3}
\end{equation}

\subsection{Index families}
\label{sec:families}

Figure~\ref{fig8} presents the indices as a function of Mg$_2$ before
(Figure~\ref{fig8}a) and after (Figure~\ref{fig8}b)
velocity-dispersion correction for all galaxy observations through the
nominal aperture ($1\farcs4\times4\arcsec$; Figures~\ref{fig8} and
\ref{fig8c} include nuclear and non-nuclear observations).  Almost all
line-strength--Mg$_2$ distributions tighten slightly, except
H$\beta$--Mg$_2$, in which the scatter increases somewhat since the
velocity dispersion corrections multiply the scatter already present.
Figure~\ref{fig8c} presents the indices as a function of velocity
dispersion after velocity-dispersion correction for the same galaxies.
In Figure~\ref{fig8}, a tail of points to lower index values is
visible for strong-lined objects in both H$\beta$ and Fe5015.  This
tail is due to residual emission-line contamination in a few objects.

\placefigure{fig8}
\placefigure{fig8c}

After correction, indices seem to fall into three general families:
(1) $\alpha$-element-like indices, including both CN indices, all
three Mg indices, Na D, and TiO$_2$, characterized by relatively
narrow, positive correlations with both Mg$_2$ and velocity
dispersion; (2) Fe-like indices, including both Ca indices, the G
band, TiO$_1$, and all Fe indices, with quite broad distributions that
are only weakly correlated with Mg$_2$ and velocity dispersion; and
(3) H$\beta$, which acts inversely to the $\alpha$-element indices,
with a relatively narrow, negative correlation with Mg$_2$ and
velocity dispersion.  Similar correlations were seen in a restricted
set of indices by Burstein et al.\ (1984), Carollo et al.\ (1993) and
J{\o}rgensen (1997).  C$_2$4668 seems to be intermediate to the
$\alpha$- and Fe-like indices, with a relatively broad, but positive
correlation with Mg$_2$ and velocity dispersion.  These groupings
probably represent element abundance families with different
nucleosynthesis histories, as discussed in Worthey (1996).

\section{Remaining Systematic Errors}
\label{sec:syst}

We now estimate the remaining systematic errors in the Lick/IDS data.
Even small systematic errors are a source of concern because indices
change only slightly over time for old stellar populations, so that
small index differences can translate to significant age differences.
For example, a systematic error in the key H$\beta$ index of only 0.05
\AA\ corresponds to a model age difference of $\sim$ 1 Gyr at 15 Gyr
(Worthey 1994).

There are two potential sources of inhomogeneities, and thus
systematic errors, in the data.  One comes from the use of two
measurement schemes, the original scheme described by Burstein et
al. (1984) (hereafter called ``eye'') and the current scheme used here
and for many stars in Paper V (called ``AUTOINDEX'').  The second
source of error comes from the presence of two separate instrumental
systems (for the first 11 indices only) --- an earlier one (called
``old'') based on standardizing to mean data for K giant standards in
Runs 3--24, and a second one (called ``new'') based on K giant
standards from all runs.  The original 11 indices published for K
giants (Faber et al.~1985) and G dwarfs (Gorgas et al.~1993) were
measured with the {\it eye} method and transformed to the {\it old}
system, whereas the {\it new} stellar data in Paper V and the galaxy
and globular data measured here were measured with AUTOINDEX and
transformed (at least initially, see below) to the {\it new} system.
We therefore consider (1) systematic differences in raw measurements
between the eye and AUTOINDEX schemes and (2) any zeropoint
differences and their uncertainty between the old and new standard
systems.  We stress that these issues exist only for the 11 original
indices; the 10 new indices added in Paper V have always been measured
using AUTOINDEX and standardized to the K giant data from all runs.

\subsection{Measurement systematics}

We begin with a comparison of the eye and AUTOINDEX schemes;
there are two principal differences between them.
\begin{enumerate}
\item Centering of feature bandpasses.  In the eye scheme, wavelength
errors were corrected by centering feature bandpasses by eye using a
reference stellar spectrum.  AUTOINDEX centers features automatically
by performing a cross-correlation of the object spectrum with a
template stellar spectrum.  These automatic centerings were then
checked visually by eye.
\item Continuum determination.  As discussed in
Section~\ref{sec:2.2b}, the eye scheme took the continuum to be
horizontal over the feature bandpass at a level $F_{C\lambda}$
measured at the midpoint of the bandpass.  In the AUTOINDEX scheme,
the continuum slopes over the feature bandpass.  The difference in
continuum shapes potentially induces small, systematic shifts in the
feature strengths.
\end{enumerate}

Figures~\ref{figb1}--\ref{figb4} investigate these potential errors by
plotting the quantity (eye$-$AUTOINDEX) for stars and galaxies
(including globular clusters) separately.  All galaxy and globular
cluster observations are plotted in Figures~\ref{figb1} and
\ref{figb2}, including off-nucleus and non-standard aperture size
observations (i.e., all observations represented in
Tables~\ref{tbl6a}--\ref{tbl6d} are including in these Figures).  All
of these are raw values with no run or velocity dispersion corrections
applied.

\placefigure{figb1}
\placefigure{figb2}
\placefigure{figb3}
\placefigure{figb4}

Figures~\ref{figb1} and \ref{figb3} plot (eye$-$AUTOINDEX) vs. eye
values.  Most of the outlying points are either M stars
(Fig.~\ref{figb3}) or very noisy galaxy spectra (Fig.~\ref{figb1}).
For either, small centering differences between the two schemes can
make large differences in the index values.  A few residual
distributions are also skewed toward negative values (e.g., Fe 5270,
Fe5335).  This probably results from the systematically better index
centering in AUTOINDEX, which results in larger index values.
However, these effects are small.

Figures~\ref{figb2} and \ref{figb4} plot (eye$-$AUTOINDEX) vs. run
number.  Run-to-run differences are seen of order $\le0.2$ \AA\ and
$\le0.010$ mag, reflecting changes in instrumental response (i.e.,
spectral shape) among runs.  (These are about half the size of the
applied run corrections).  However, large-scale, systematic trends
that affect a large fraction of the data are at most half this size.

Of concern from the standpoint of systematic errors is any {\it global
shift} or {\it tilt} between the two measuring schemes.  Mean
differences between eye and AUTOINDEX are summarized in
Table~\ref{tblb1} for stars and galaxies separately.  Except for
CN$_1$, global shifts are generally very small, $\le0.04$ \AA\ and
$\le0.003$ mag.  CN$_1$ shows an offset of 0.005 mag for stars, plus a
tilt of comparable size (see Fig.~\ref{figb3}).  Both effects were
mentioned in Paper V, but neither seems to be present for galaxies and
globular clusters (cf.\ Fig.~\ref{figb1}).  Neither the origin of
these trends nor the difference between stars and galaxies are
understood.

\placetable{tblb1}

Summarizing the information in Figures~\ref{figb1}--\ref{figb4} and
Table~\ref{tblb1}, we conclude that large-scale, systematic
differences between the eye and AUTOINDEX measuring schemes are
generally $\le0.05$ \AA\ and $\le0.003$ mag, with the exception of
CN$_1$, for which the differences are twice as large.

We turn now to differences between the ``old'' and ``new'' standard
systems.  Recall that the standard system for the 11 original indices
(here called the ``old'' system) was standardized to the K giant
standards in Runs 3 -- 24, about one-third of the data.  In hindsight,
we see that these early runs were atypical in some indices and that
the standard system is therefore slightly ``off'' with respect to the
whole data.  Rather than change zeropoints now, since many data have
been published and fitting-functions derived from them (Gorgas et
al.~1993; Paper V), we compute zeropoint corrections needed to
transform AUTOINDEX plus its new system of run corrections to the old,
published system.  The adopted corrections, based on the 9 K giant
standard stars, are given in the last column of Table~\ref{tblb1};
they are applied to all the galaxy and globular data in this paper.
For most indices, the corrections are quite small, a few hundredths of
an \AA\ or a few thousandths of a magnitude.  The significant
exception is the G-band, for which the offset is 0.21 \AA.  A
previous, similar analysis in Paper V yielded the corrections shown in
the second-to-last column.  These shifts were used to correct the new
stellar data in Paper V.  The differences between the two sets of
corrections are again at most a few hundredths of an \AA\ or a few
thousandths of a magnitude.  These are small to negligible in the
context of old stellar-populations.  The differences between the Paper
V and present corrections are a measure of the irreducible zeropoint
uncertainties inherent in the published Lick/IDS system.

\subsection{Mg$_2$:  Comparison with Seven Samurai}

Finally, we examine the Mg$_2$ values presented here with respect to
those of the Seven Samurai (\cite{Davies87}, \cite{Faber89}).  Davies
et al.\ (1987) used a combined Mg$_2$ index that weighted
contributions from Mg$_2$ and Mg$_1$, both measured using the eye
scheme.  The resulting Mg$_2$ index is hereafter called $\langle \rm
Mg_2\rangle$ to distinguish it from the Lick/IDS index Mg$_2$.  We
reproduce Equations 2 and 3 of Davies et al.\ here:
\begin{equation}
\rm Mg_2^{\prime}=0.03 + 2.10\, Mg_1 - 62\, Mg_1^4,\label{eqmg2p}
\end{equation}
\begin{equation}
\rm \langle{Mg_2}\rangle = 0.6\, Mg_2 + 0.4\, Mg_2^{\prime}.\label{eqmg2a}
\end{equation}
We have recomputed $\langle \rm Mg_2\rangle$ using the AUTOINDEX
measurements for all galaxies in common between the two samples (Seven
Samurai and that presented here).  After removing the aperture
correction to the Seven Samurai measurements (Equation 4 of Davies et
al.), we compare the results in Figure~\ref{figb6}.  The mean
difference (Seven Samurai$-$AUTOINDEX) is $+0.003$ mag, with a
standard deviation of 0.010.  This difference is close to what one
would expect from comparing of the eye and AUTOINDEX schemes for
Mg$_2$ in Table~\ref{tblb1}.  The dispersion is also expected from a
close examination of Figure~\ref{figb1}.  We recommend that those
interested in using $\langle \rm Mg_2\rangle$ for Lick galaxies
recompute this index from the values of Mg$_1$ and Mg$_2$ given here.

\placefigure{figb6}

\section{Final Absorption-Line Indices}
\label{sec:2.5}

Table~\ref{tbl6a} presents final mean velocity-dispersion-corrected
indices, rms errors and total goodnesses $\sqrt{N_{obs}}\langle
G\rangle$ for all galaxy nuclei observed through the nominal slit
width and length ($1\farcs4\times4\arcsec$).

\placetable{tbl6a}

Table~\ref{tbl6b} presents similar data for nearly all globular
clusters in the sample.  Globular clusters have stellar run
corrections applied to all indices.  The values in Tables~\ref{tbl6a}
and \ref{tbl6b} supersede all previously published Lick/IDS galaxy and
globular cluster index strengths.  Galactic globular clusters were
scanned over the cluster through a $1\farcs4 \times 16\arcsec$ slit to
synthesize a 66\arcsec$\times$66\arcsec\ aperture (cf. Burstein et
al. 1984).  Entries marked ``O'' are off-center observations through a
similarly scanned aperture displaced 35\arcsec\ away from the main
aperture.

\placetable{tbl6b}

Table~\ref{tbl6c} presents data for galaxies observed through the
nominal aperture of $1\farcs4\times4\arcsec$ but off the nucleus.  The
offset from the nucleus in arcseconds is marked next to the galaxy
name.  See the notes for details.

\placetable{tbl6c}

Table~\ref{tbl6d} presents data for observations through non-standard
apertures.  These were mostly off-nuclear measurements of bright
galaxies, plus a few wide-slit nuclear observations of small galaxies
and two globular clusters (the M31 globulars V29 and V92).  The slit
was widened to increase signal-to-noise.  The offset from the nucleus
(typically in arcseconds) is marked next to the galaxy name, if
applicable.  See the notes for details.  Column 2 lists the aperture
dimensions; entries marked ``scan'' in Column 2 were spatially scanned
over a $16\arcsec\times16\arcsec$ area through a $1\farcs4 \times
16\arcsec$ slit.  Standard run corrections were applied as described
in Section~\ref{sec:2.2b}.  K giant standard stars observed through
wide slits were confirmed to have run corrections consistent with
those observed through the nominal slitwidth.

\placetable{tbl6a}

In order to bring the wide-slit galaxy observations in
Table~\ref{tbl6d} onto the Lick/IDS system, a correction for slitwidth
broadening was made that was similar to the velocity dispersion
correction.  Figure~\ref{fig9} shows a plot of the K giant standard
star indices measured through wide slits as a function of slit width
(compare to Figure~\ref{fig7}).  Observations through $1\farcs8$- and
$2\farcs2$- slits were judged usable without need for correction.  For
observations through the $3\farcs4$-, $5\farcs4$-, and $7\farcs4$-wide
slits, the median values of the K giant ratios of mean index strength
through the nominal slit to the wide-slit index strengths were used to
correct the index values and raw errors.  These multiplicative
corrections are listed in Table~\ref{tbl7}.  The strengths of Ca4227,
Ca4455, Fe4531, H$\beta$, Fe5015, Fe5335, Fe5406, Fe5709, and Fe5782
were all judged to be unusable for observations through the
$7\farcs4$-wide slit due to the large dispersion in the K giant ratios
of Figure~\ref{fig9}.  These indices are not listed for this aperture
in Table~\ref{tbl6d}.

\placefigure{fig9}
\placetable{tbl7}

Some index measurements are missing from
Tables~\ref{tbl6a}--\ref{tbl6d}.  There are five possible reasons: (1)
The spectral coverage of the IDS system was not consistent throughout
all runs, and the CN indices or TiO indices may not have been observed
(this is more likely for galaxies observed only once).  (2) Ephemeral
features caused by bubbles in the immersion oil of the photomultiplier
chain may have contaminated certain index measurements.  (3) The
systemic velocity of the galaxy may have moved the reddest indices
(TiO$_1$ and TiO$_2$) out of the spectral range of the IDS system.
(4) Intrinsic emission such as H$\beta$ or [\ion{O}{3}] $\lambda5007$
in the galaxy may have contaminated a central bandpass or sideband.
We have culled the most obvious examples of emission contamination,
but subtle contamination remains.  Users of these data should be aware
of this.  (5) Poorly-subtracted night-sky lines contaminated certain
indices.  Table~\ref{tbl8} presents a list of indices possibly
affected by residual contamination from poor night-sky subtraction.

\placetable{tbl8}

\section{Summary}
\label{sec:2.conc}

This paper presents the complete database of Lick/IDS absorption-line
index strengths for galaxies and globular clusters.  This database
supersedes all previously published Lick/IDS data on these objects.
The Lick/IDS galaxy data are among the largest collection of
homogeneous absorption-line strengths for stars and galaxies currently
available.

We have reviewed the measurement of Lick/IDS indices from IDS spectra
and characterized the errors.  The level of remaining systematic
uncertainties is discussed.  We also present for the first time the
correction of Lick/IDS absorption-line strengths for velocity
dispersion.  Such a correction is a crucial step to compare Lick/IDS
galaxy absorption-line strengths to models of stellar populations
based on the Lick/IDS stellar library (\cite{Worthey94}).

In a subsequent paper, we will present an analysis of a subset of
these data using stellar population models in an attempt to derive
stellar population ages, metallicities, and relative element
abundances of the nuclei of early-type galaxies.

\acknowledgments

The authors would like to thank their previous collaborators on this
project, particularly C.~Dalle Ore; the directors, telescope
operators, and staff of Mount Hamilton, Lick Observatory; J.~Wampler
and L.~Robinson for their development of the Image Dissector Scanner;
and the referee, J.~Rose, for helpful comments.  This work was
supported by NSF grants AST 76-08258, 82-11551, 87-02899, and 95-29008
to SMF and AST 90-16930 to DB; by an ASU Faculty Grant-in-Aid to DB;
by the WFPC Investigation Definition Team contract NAS 5-1661; NASA
grant HF-1066.01-94A to GW from the Space Telescope Science Institute,
which is operated by the Association of Universities for Research in
Astronomy, Inc., under NASA contract NAS5-26555; and by a Flintridge
Foundation Fellowship and by a Starr Fellowship to SCT.

\clearpage


\clearpage

\clearpage

\begin{figure}
\plotone{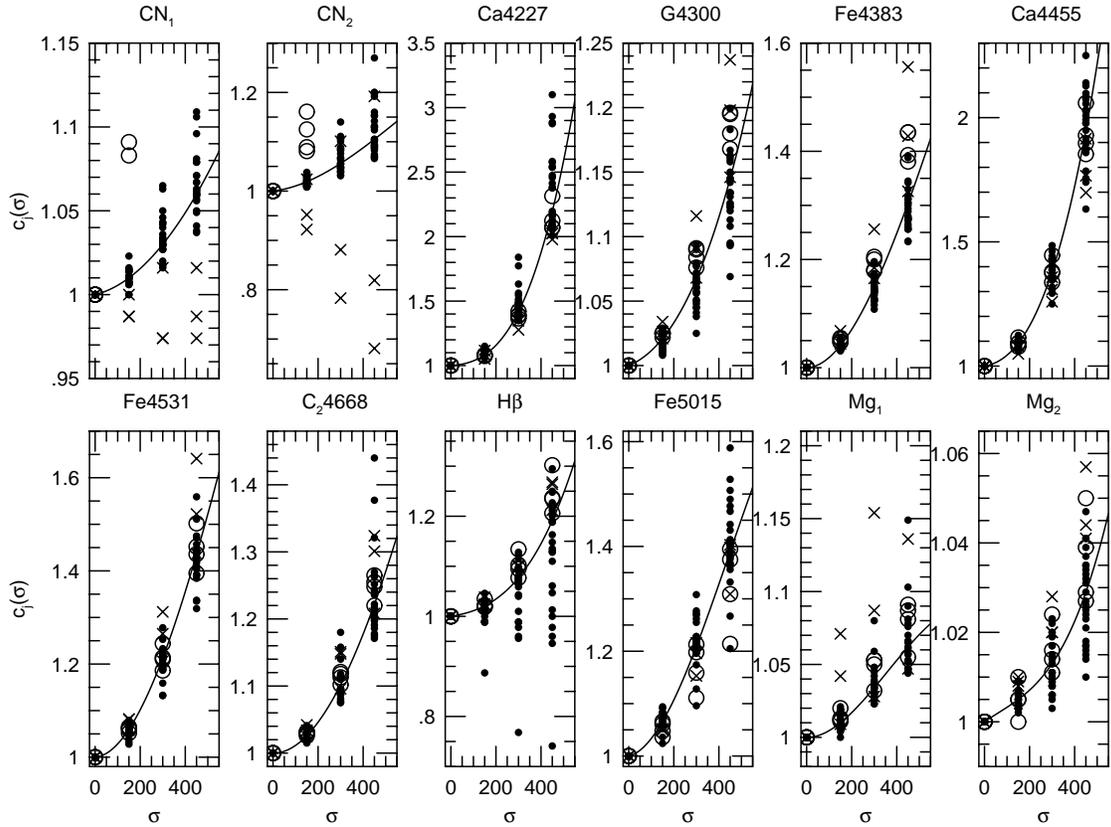}
\caption{Multiplicative index correction, $C_j$, as a function of
velocity dispersion (in \kms) for all 21 Lick/IDS indices.  The
corrections have been determined by measuring various stellar spectra
convolved to different broadenings.  Symbols represent K giant
standard stars (small dots), G dwarfs (crosses), and M32 (multiple
observations; open circles).  The quantity shown is
index(0)/index($\sigma$).\label{fig7}}
\end{figure}

\clearpage

\begin{figure}
\plotone{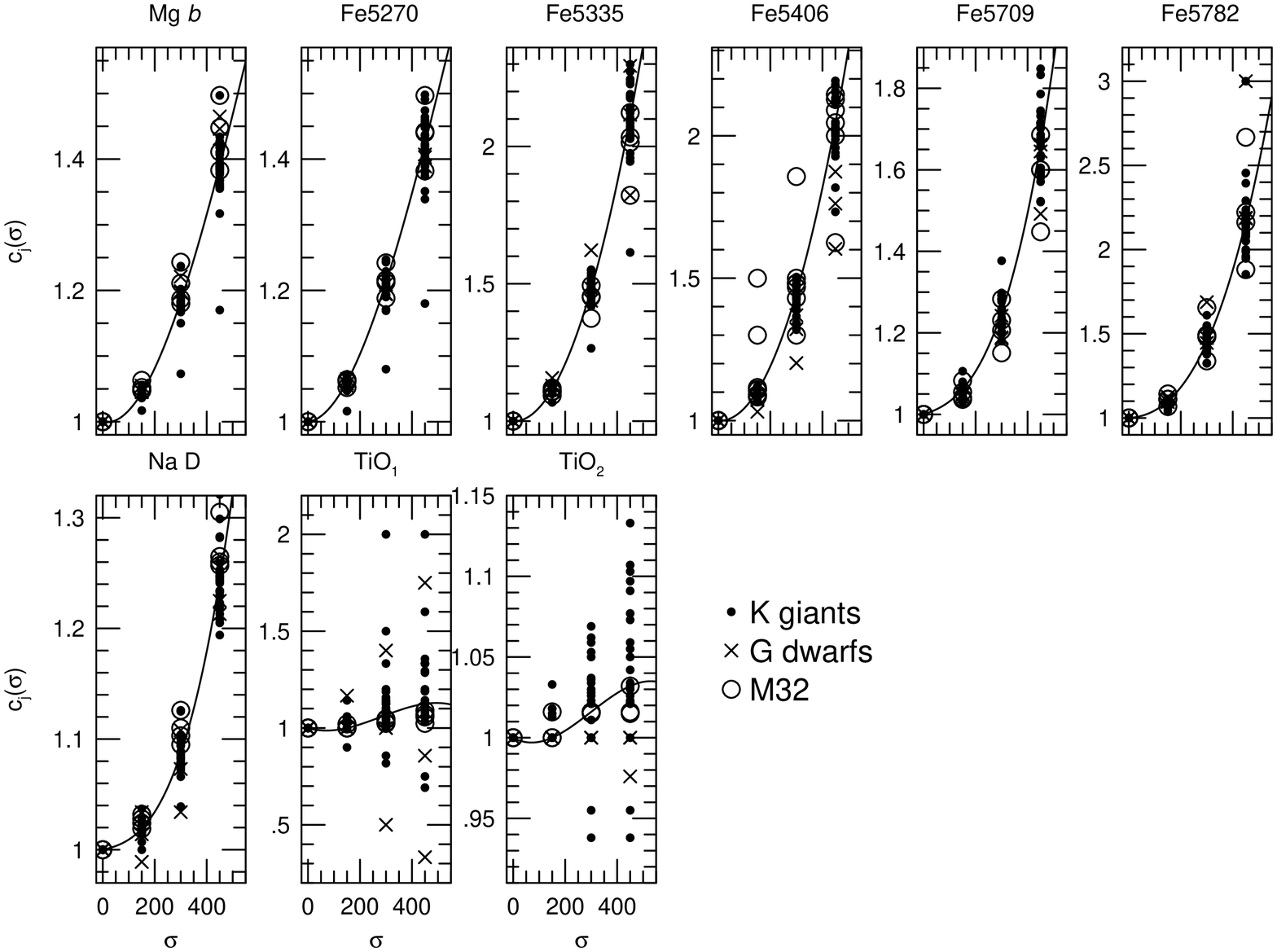}
\figurenum{3}
\caption{Continued.}
\end{figure}

\clearpage

\begin{figure}
\plotone{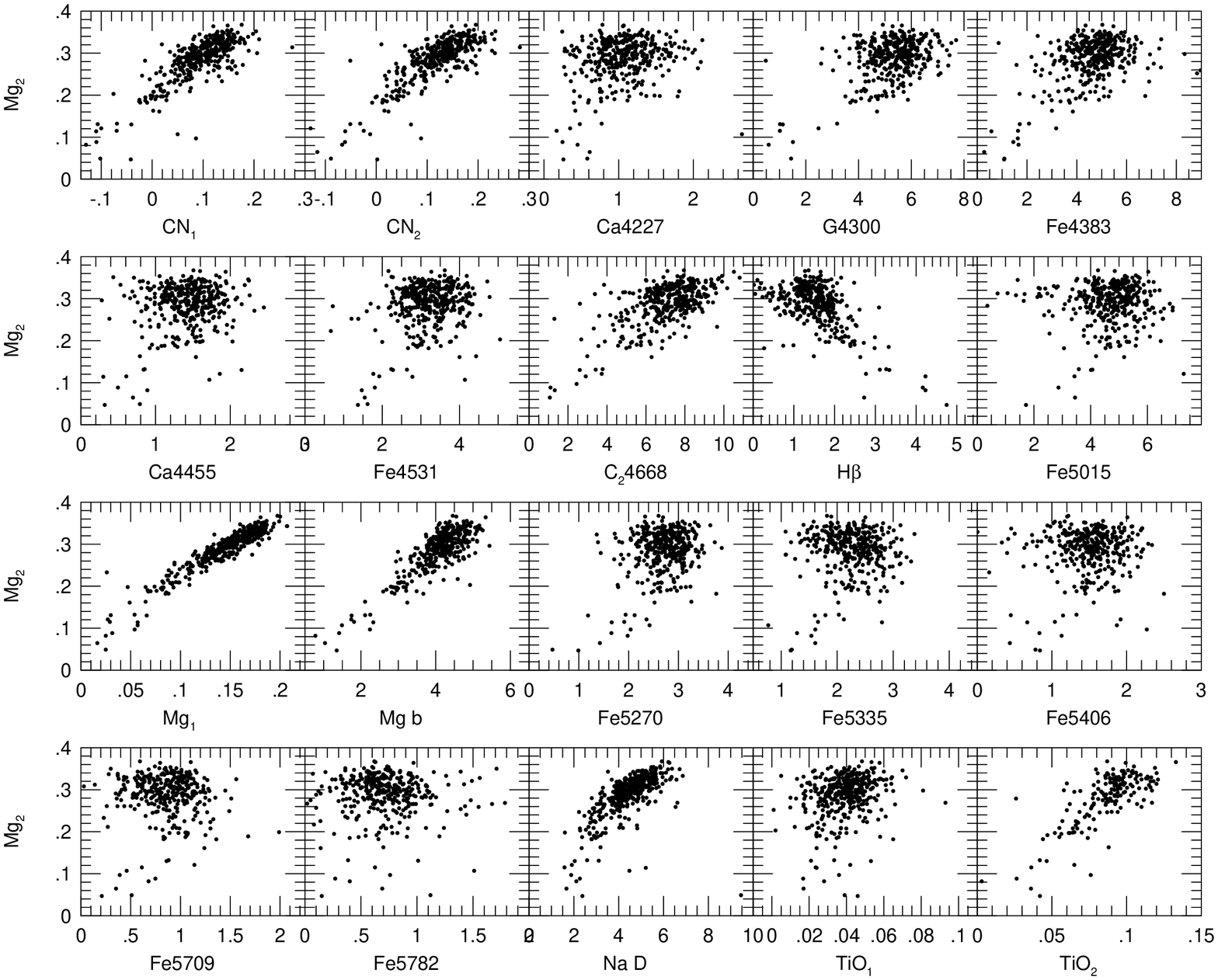}
\caption{Index strengths as a function of Mg$_2$ (in mag) for all
21 Lick/IDS indices for all galaxy observations (nuclear and
off-nuclear) through the standard $1\farcs4\times4\arcsec$ aperture.
(a) Before velocity-dispersion correction.\label{fig8}}
\end{figure}

\clearpage

\begin{figure}
\plotone{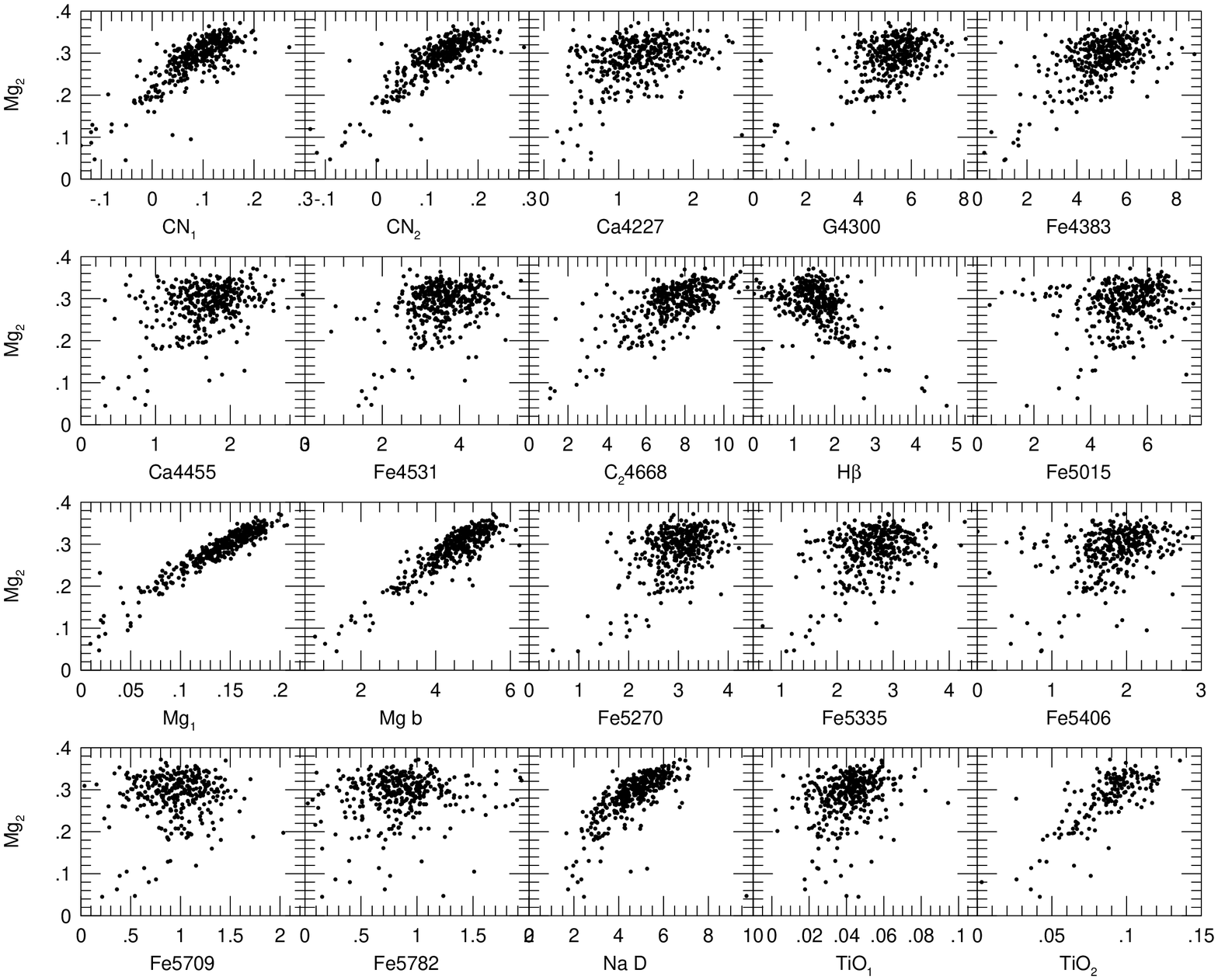}
\figurenum{4}
\caption{(b) After velocity-dispersion correction.}
\end{figure}

\clearpage

\begin{figure}
\plotone{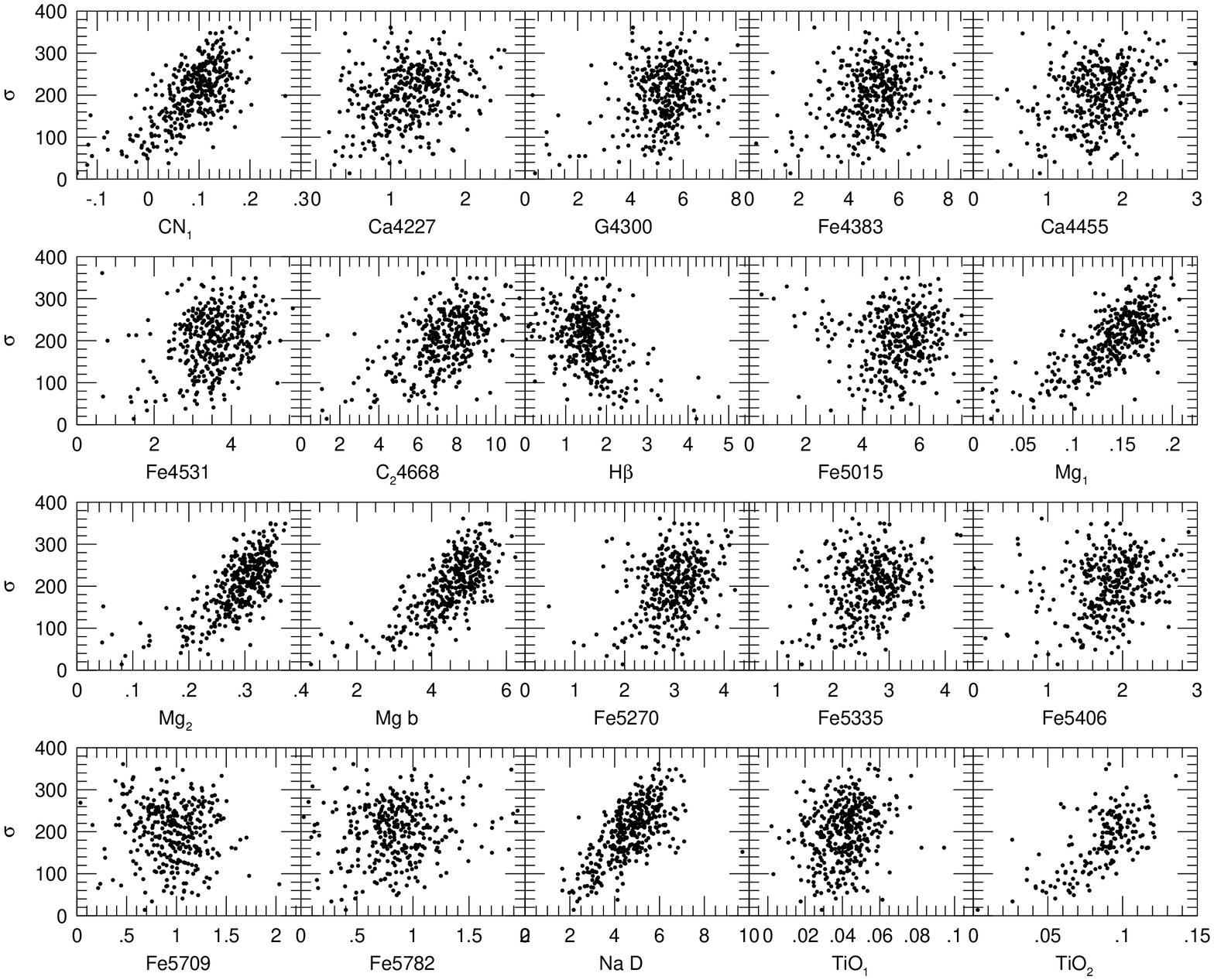}
\caption{Index strengths as a function of velocity dispersion (in
\kms) for 20 Lick/IDS indices (all except CN$_2$, which reproduces the
behavior of CN$_1$ very closely) for all galaxy observations (nuclear
and off-nuclear) through the standard $1\farcs4\times4\arcsec$
aperture, after velocity-dispersion correction.\label{fig8c}}
\end{figure}

\clearpage

\begin{figure}
\plotone{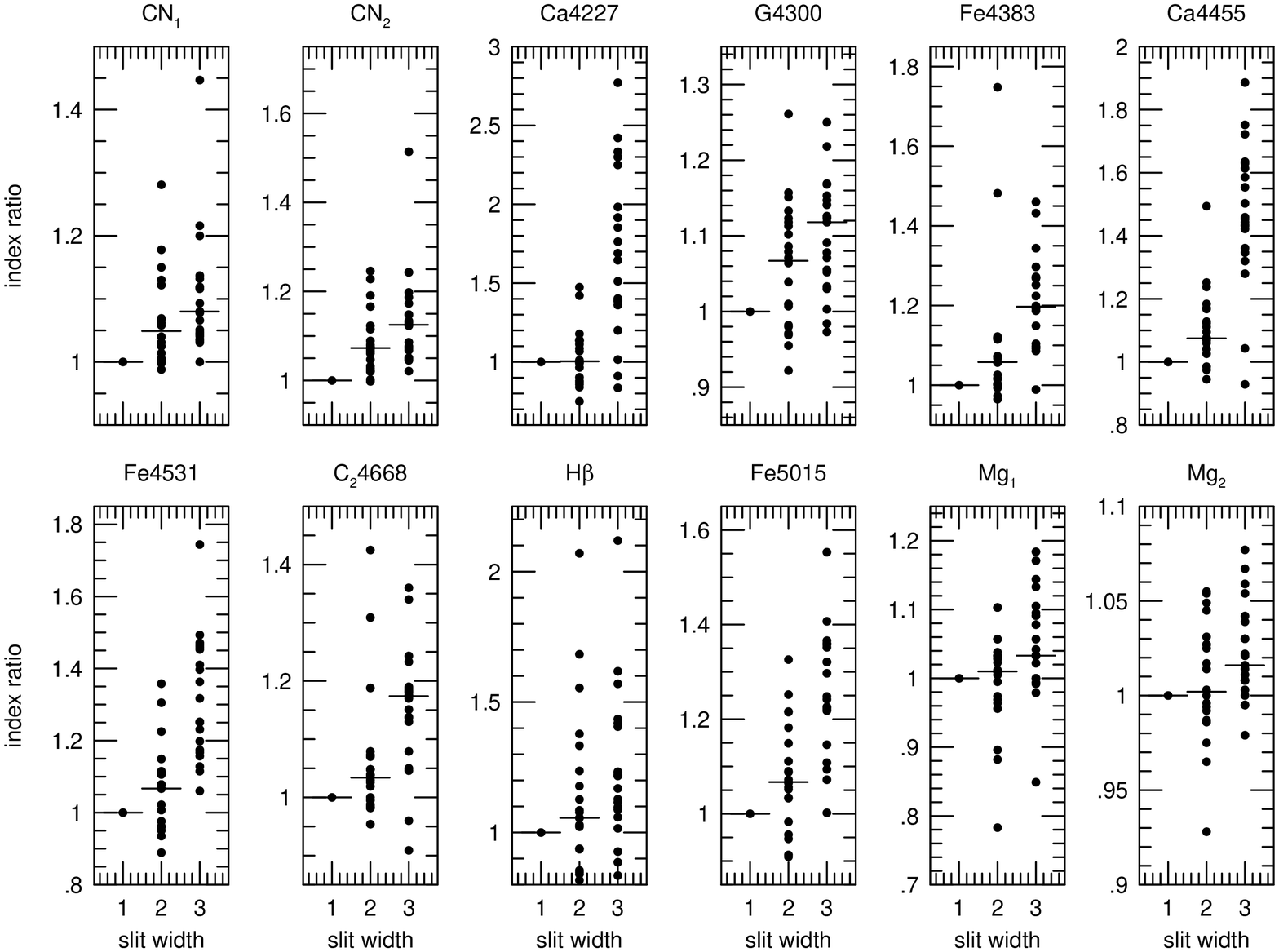}
\caption{Multiplicative index corrections as a function of
slit\-width (1=nominal slit\-width, $1\farcs 4$; 2=$3\farcs4$;
3=$7\farcs4$) for all 21 Lick/IDS indices, for K giant standard stars.
The vertical axis is index($1\farcs4$)/index(slitwidth).  Horizontal
lines are the corrections applied at each slitwidth.  Missing
horizontal lines denote indices deemed too uncertain at $7\farcs4$ to
be useful.\label{fig9}}
\end{figure}

\clearpage

\begin{figure}
\plotone{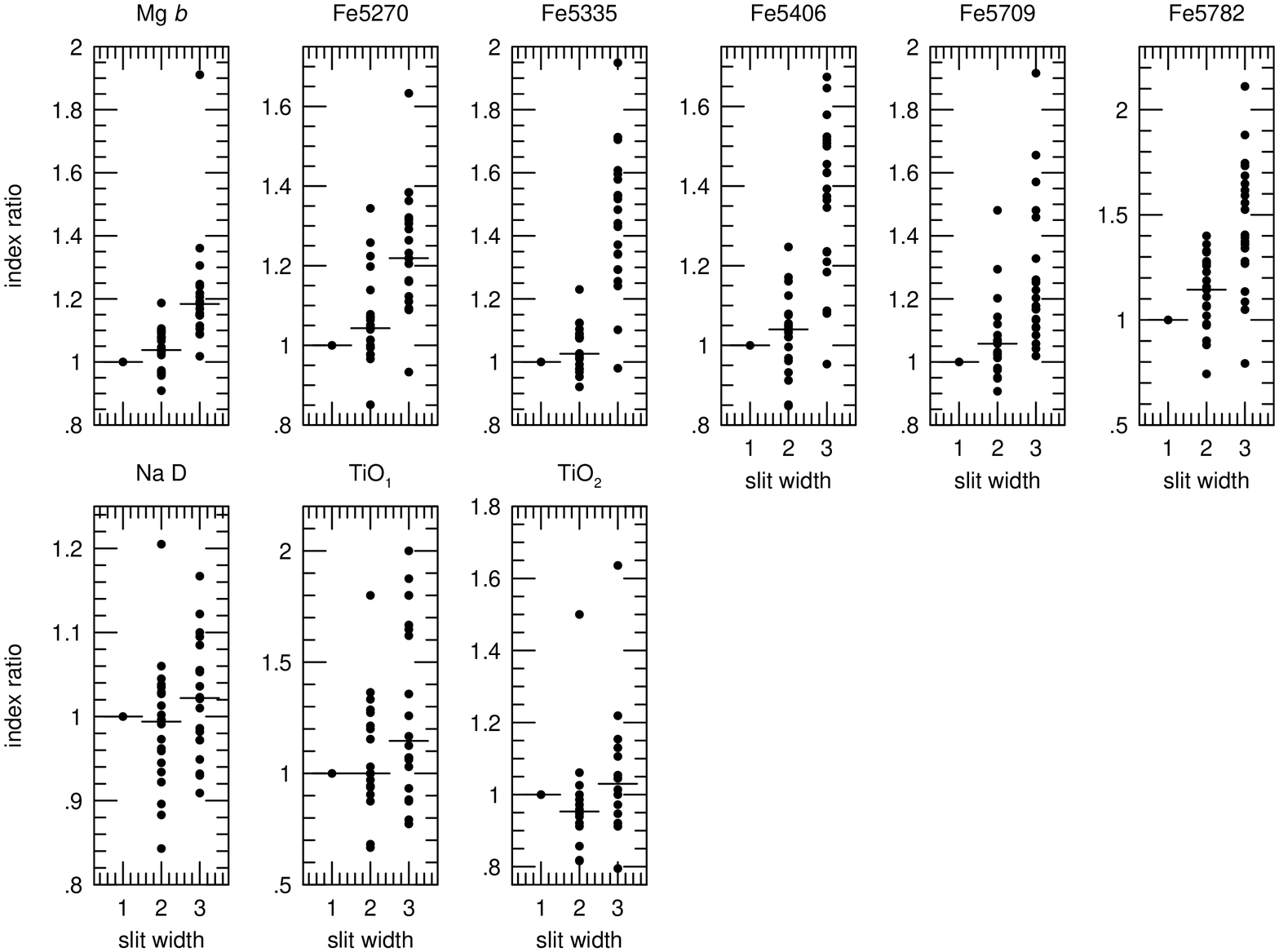}
\figurenum{6}
\caption{Continued.}
\end{figure}

\clearpage

\begin{figure}
\plotone{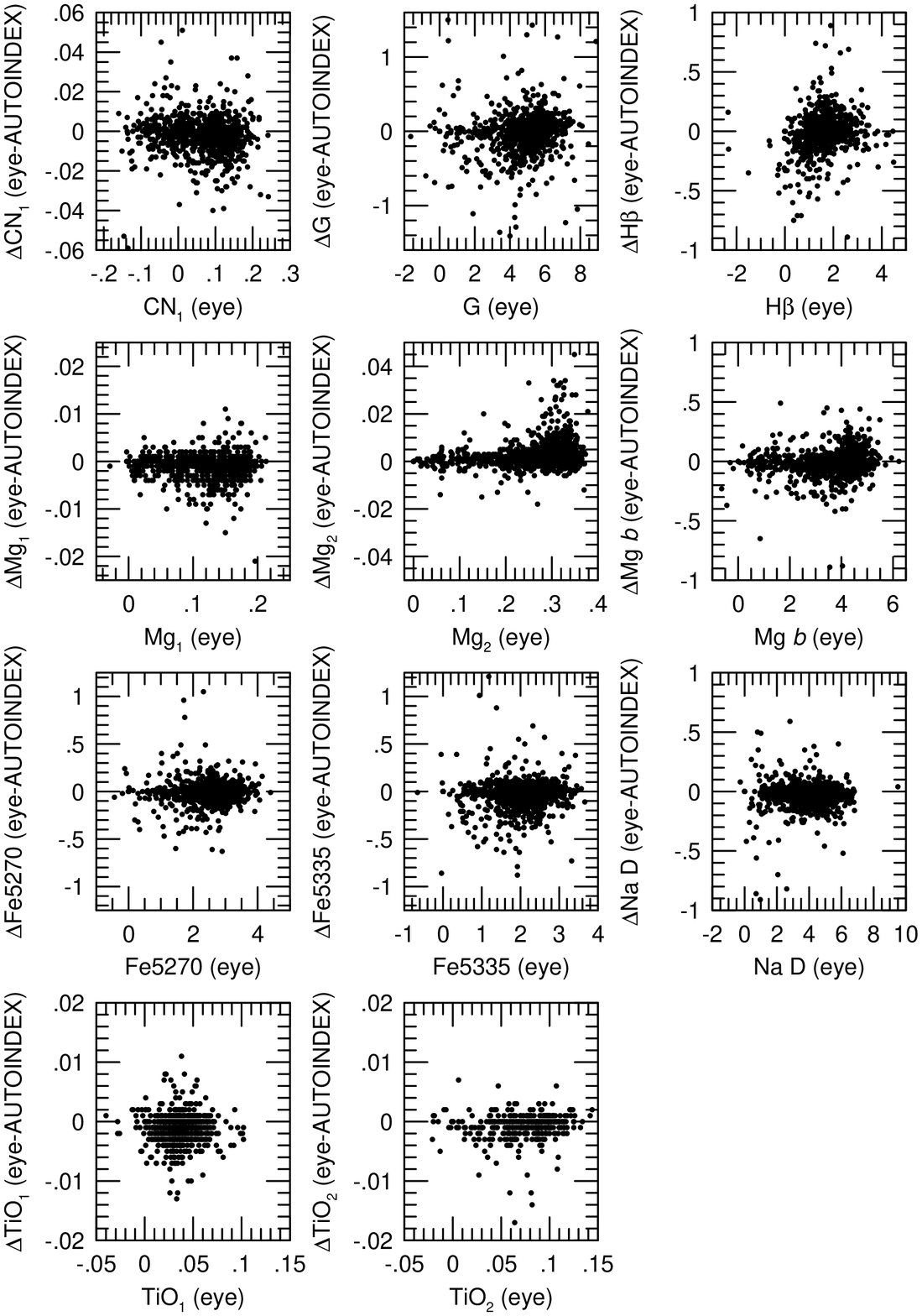}
\caption{IDS measurement scheme differences, eye$-$AUTOINDEX, for
all galaxy and globular cluster observations (including off-nuclear
and wide-slit observations), as a function of eye measurements.  Run
and velocity-dispersion corrections have not been
applied.\label{figb1}}
\end{figure}

\clearpage

\begin{figure}
\plotone{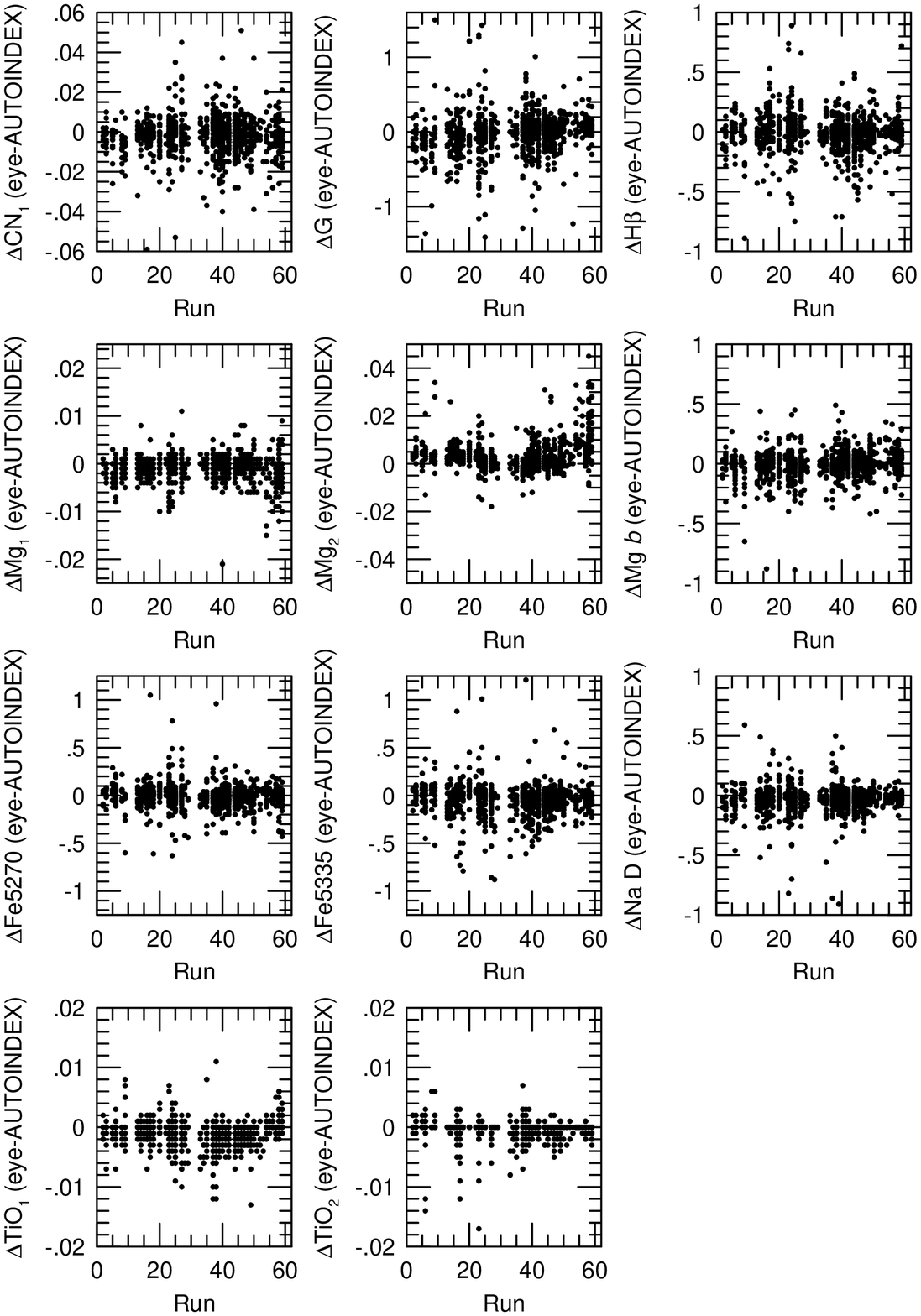}
\caption{IDS measurement scheme differences, eye$-$AUTOINDEX, for
all galaxy and globular cluster observations (including off-nuclear
and wide-slit observations), as a function of IDS run number.  Run
corrections have not been applied.\label{figb2}}
\end{figure}

\clearpage

\begin{figure}
\plotone{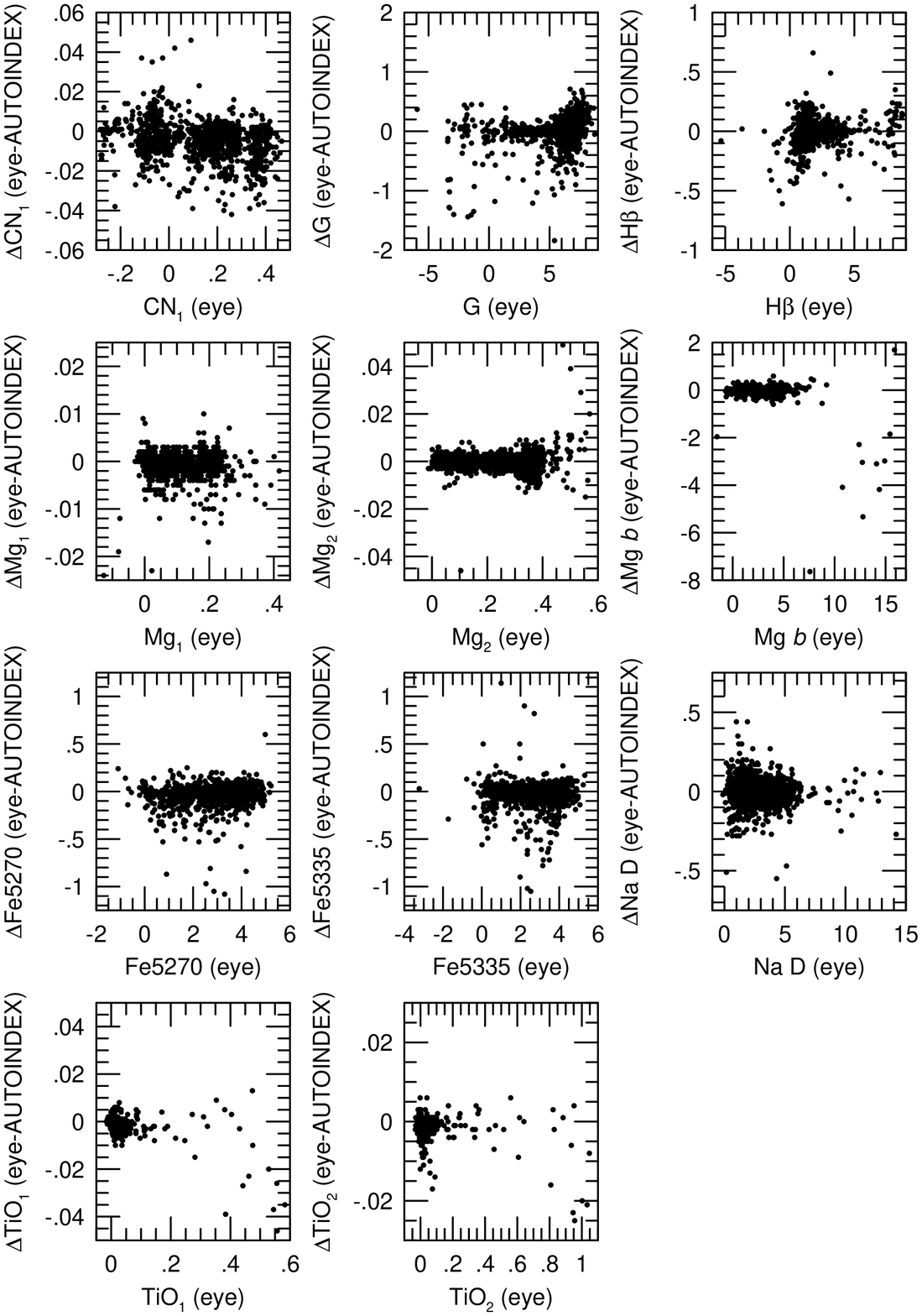}
\caption{IDS measurement scheme differences, eye$-$AUTOINDEX, for
all stars in Paper V, as a function of eye measurements.  Run
corrections have not been applied.\label{figb3}}
\end{figure}

\clearpage

\begin{figure}
\plotone{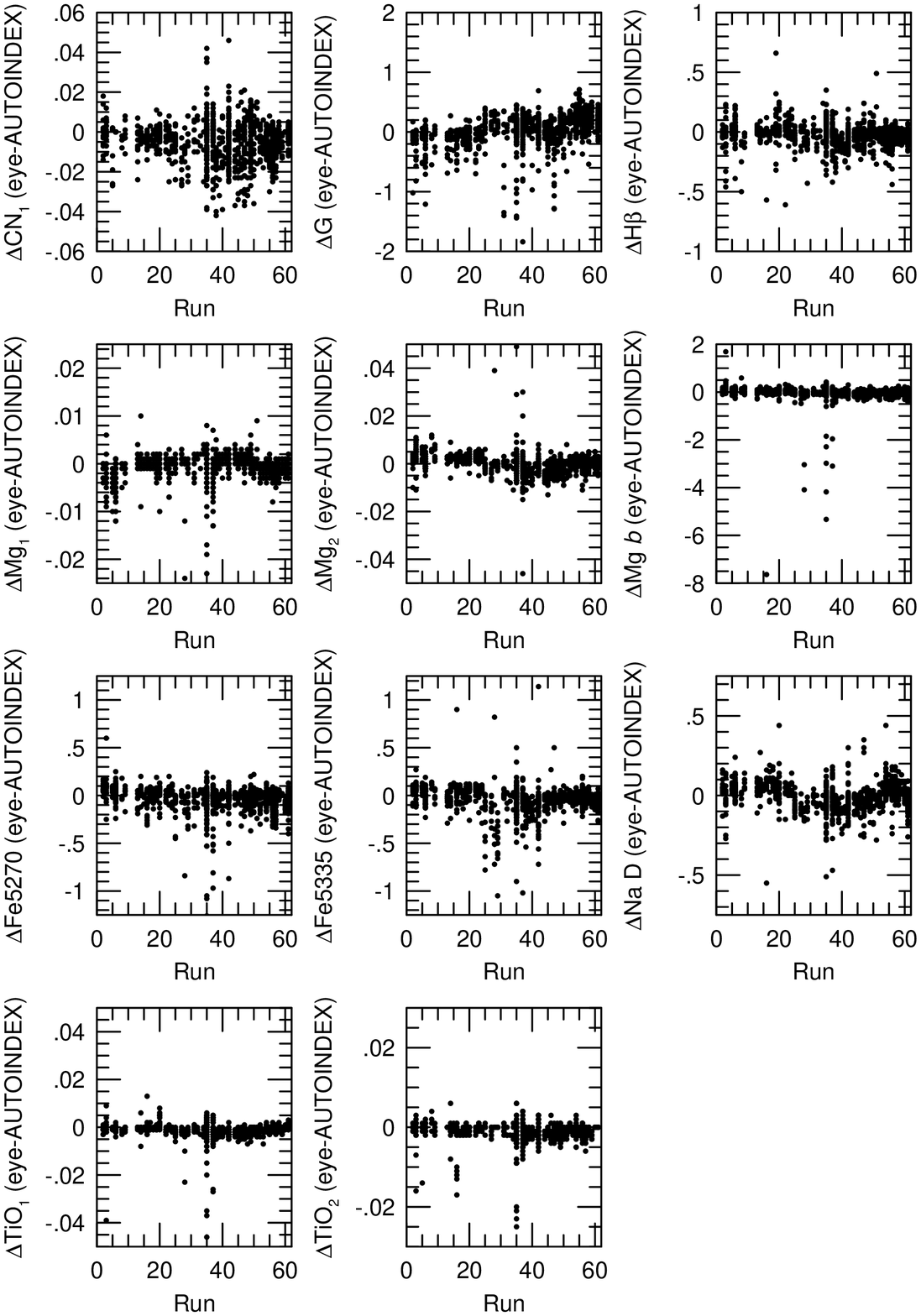}
\caption{IDS measurement scheme differences, eye$-$AUTOINDEX, for
all stars in Paper V, as a function of IDS run number.  Run
corrections have not been applied.\label{figb4}}
\end{figure}

\clearpage

\begin{figure}
\plotone{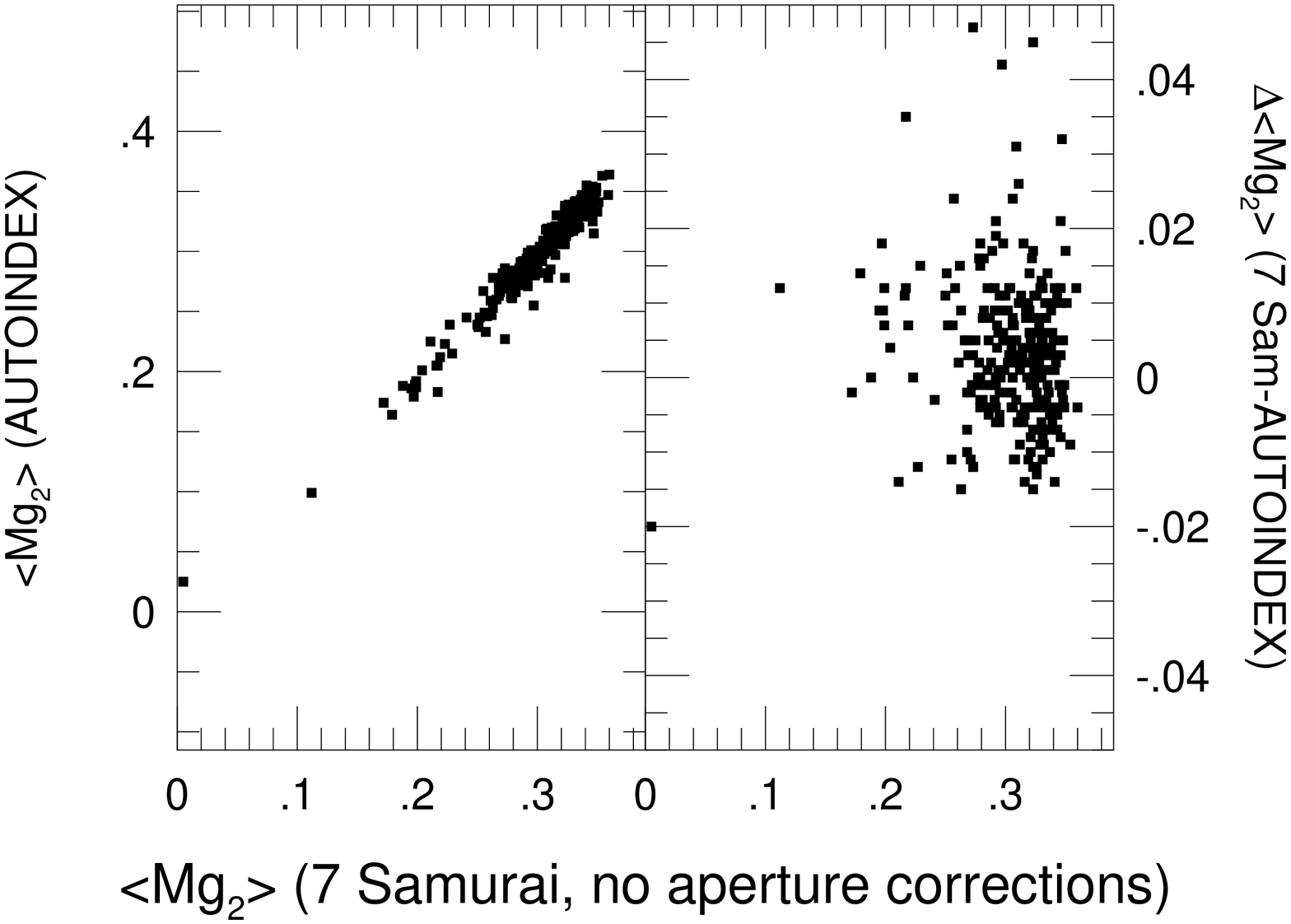}
\caption{Comparison of IDS and Seven Samurai measurements of
$\langle \rm Mg_2\rangle$.  Data for Seven Samurai measurements are
taken from Davies et al.\ (1987), without aperture corrections to
Coma.\label{figb6}}
\end{figure}

\clearpage


\begin{deluxetable}{lccl}
\scriptsize
\tablecaption{Published IDS Data\label{tbl1}}
\tablewidth{0pt}
\tablehead{\colhead{Paper}&\colhead{No.\ of Indices}&\colhead{Method}&
\colhead{Run corrections}}
\startdata
\cutinhead{(a) Galaxies and Globular Clusters}
Burstein et al.\ (1984) globulars& 11 & E & all indices\nl
Davies et al.\ (1987) & $\langle{\rm Mg_2}\rangle$\tablenotemark{a} &
E & yes\nl
Burstein et al.\ (1988) & $\langle{\rm
Mg_2}\rangle$\tablenotemark{a}\tablenotemark{b} & E & yes\nl
Worthey, Faber, \& Gonz{\'a}lez (1992) & Mg$_2$, Fe5270, Fe5335 & A &
Mg$_2$ only\nl
This paper & & & \nl
\qquad Galaxies & 21 & A & molecular bands only\nl
\qquad Globulars, low-velocity galaxies\tablenotemark{c} & 21 & A &
all indices\nl
\nl
\cutinhead{(b) Stars}
Faber et al.\ (1985) K giants& 11 & E & all indices \nl
Burstein et al.\ (1986) & Fe5270, Fe5335 & E & all indices\nl
Gorgas et al.\ (1993) G dwarfs& 11 & E & all indices\nl
Worthey et al.\ (1994) & & & \nl
\qquad Prev.\ published K giants, G dwarfs & 11 & E & all indices \nl
 & +10 & A & all indices \nl
\qquad All other stars & 21 & A & all indices \nl
\enddata
\tablenotetext{a}{The $\langle{\rm Mg_2}\rangle$ index is a weighted
mean of Mg$_1$ and Mg$_2$.  See Section 6.2.}
\tablenotetext{b}{There is an error in the $\langle{\rm Mg_2}\rangle$ index
for NGC 3115 in Table 3 of Burstein et al.\ (1988); the correct value
is $\langle{\rm Mg_2}\rangle=0.330$.  Note that the $\langle{\rm Mg_2}\rangle$
values in Table 3 of Burstein et al.\ (1988) are from Davies et al.\
(1987), without the aperture correction of Davies et al.}
\tablenotetext{c}{Galaxies with $cz < 300$\ \kms.}
\tablecomments{Columns:\\
(1) Reference\\
(2) Number of indices published: 11=original 11
Lick/IDS indices of Burstein et al.\ (1984); 21=all Lick/IDS indices
(see Table 2); +10=new indices presented in Worthey et al.\ (1994);
$\langle{\rm Mg_2}\rangle$=``average'' Mg$_2$ index described in Davies
et al.\ (1987; cf.\ Section 6).\\
(3) Index measurement method: E=``eye'' [see Burstein
et al.\ (1984)]; A=AUTOINDEX (see text).\\
(4) Run corrections are determined by zeropointing K giant standard
star observations to the standard system determined by the same nine
standard stars (see Faber et al.\ 1985). For further discussion of the
system, see Section 6.1.}
\end{deluxetable}

\clearpage

\begin{deluxetable}{llrrclll} 
\scriptsize
\tablewidth{0pt}
\tablecaption{Index Definitions\label{tbl2}}
\tablehead{\colhead{$j$}&\colhead{Name}&\colhead{Index Bandpass}&
\colhead{Pseudocontinua}&\colhead{Units}&
\colhead{Measures\tablenotemark{a}}&\colhead{Error\tablenotemark{b}}&
\colhead{Notes}}
\startdata
01&CN$_1$&4142.125-4177.125&4080.125-4117.625&mag&C,N,(O)&0.018&1,2\nl
&&&4244.125-4284.125&&&\nl    
02&CN$_2$&4142.125-4177.125&4083.875-4096.375&mag&C,N,(O)&0.019&1,2\nl  
&&&4244.125-4284.125&&&\nl   
03&Ca4227&4222.250-4234.750&4211.000-4219.750&\AA&Ca,(C)&0.25&1\nl    
&&&4241.000-4251.000&&&\nl    
04&G4300&4281.375-4316.375&4266.375-4282.625&\AA&C,(O)&0.33&1\nl     
&&&4318.875-4335.125&&&\nl     
05&Fe4383&4369.125-4420.375&4359.125-4370.375&\AA&Fe,C,(Mg)&0.46&1\nl     
&&&4442.875-4455.375&&&\nl     
06&Ca4455&4452.125-4474.625&4445.875-4454.625&\AA&(Fe),(C),Cr&0.22&1\nl     
&&&4477.125-4492.125&&&\nl     
07&Fe4531&4514.250-4559.250&4504.250-4514.250&\AA&Ti,(Si)&0.37&1\nl     
&&&4560.500-4579.250&&&\nl     
08&C$_2$4668&4634.000-4720.250&4611.500-4630.250&\AA&C,(O),(Si)&0.57&1,3\nl
&&&4742.750-4756.500&&&\nl     
09&H$\beta$&4847.875-4876.625&4827.875-4847.875&\AA&H$\beta$,(Mg)&0.19&\nl
&&&4876.625-4891.625&&&\nl 
10&Fe5015&4977.750-5054.000&4946.500-4977.750&\AA&(Mg),Ti,Fe&0.41&\nl     
&&&5054.000-5065.250&&&\nl     
11&Mg$_1$&5069.125-5134.125&4895.125-4957.625&mag&C,Mg,(O),(Fe)&0.006&3\nl
&&&5301.125-5366.125&&&\nl
12&Mg$_2$&5154.125-5196.625&4895.125-4957.625&mag& Mg,C,(Fe),(O)&0.007&\nl
&&&5301.125-5366.125&&&\nl
13&Mg$b$&5160.125-5192.625&5142.625-5161.375&\AA&Mg,(C),(Cr)&0.20&\nl
&&&5191.375-5206.375&&&\nl
14&Fe5270&5245.650-5285.650&5233.150-5248.150&\AA&Fe,C,(Mg)&0.24&\nl
&&&5285.650-5318.150&&&\nl
15&Fe5335&5312.125-5352.125&5304.625-5315.875&\AA&Fe,(C),(Mg),Cr&0.22&\nl 
&&&5353.375-5363.375&&&\nl
16&Fe5406&5387.500-5415.000&5376.250-5387.500&\AA&Fe&0.18&\nl
&&&5415.000-5425.000&&&\nl
17&Fe5709&5696.625-5720.375&5672.875-5696.625&\AA&(C),Fe&0.16&1\nl
&&&5722.875-5736.625&&&\nl
18&Fe5782&5776.625-5796.625&5765.375-5775.375&\AA&Cr&0.19&1\nl
&&&5797.875-5811.625&&&\nl
19&Na D&5876.875-5909.375&5860.625-5875.625&\AA&Na,C,(Mg)&0.21&1\nl
&&&5922.125-5948.125&&&\nl
20&TiO$_1$&5936.625-5994.125&5816.625-5849.125&mag&C&0.006&1,4\nl 
&&&6038.625-6103.625&&&\nl
21&TiO$_2$&6189.625-6272.125&6066.625-6141.625&mag&C,V,Sc&0.005&1,4\nl
&&&6372.625-6415.125&&&\nl
\enddata
\tablenotetext{a}{Dominant species; species in parentheses control
index in a negative sense (index weakens as abundance grows).  See
Tripicco \& Bell (1995) and Worthey (1996).}
\tablenotetext{b}{Standard star error.  See text.}
\tablecomments{\\
(1) Wavelength definition has been refined. See text.\\
(2) C, N are dominant as CN.\\
(3) C is dominant as C$_2$.\\
(4) TiO appears at M0 and cooler.}
\end{deluxetable}

\clearpage

\begin{deluxetable}{llccc}
\scriptsize
\tablecaption{Lick/IDS Error Rescalings\tablenotemark{a}\label{tbl3}}
\tablewidth{0pt}
\tablehead{
\colhead{} & \colhead{} & \colhead{G93} & \colhead{IDS-IDS} &
\colhead{Adopted} \\
\colhead{$j$} & \colhead{Name} & \colhead{rescaling} &
\colhead{rescaling} & \colhead{rescaling}}
\tablecolumns{5}
\startdata
01& CN1    & \nodata& \nodata& 0.92\nl
02& CN2    & \nodata& \nodata& 0.92\nl
03& Ca4227 & \nodata& \nodata& 0.92\nl
04& G4300  & \nodata& \nodata& 0.92\nl
05& Fe4383 & \nodata& 1.11& 1.11\nl
06& Ca4455 & \nodata& 0.87& 0.87\nl
07& Fe4531 & \nodata& 0.90& 0.90\nl
08& C$_2$4668 & \nodata& \nodata& 0.92\nl
09& H$\beta$& 0.95& \nodata& 0.95\nl
10& Fe5015 & 1.05& \nodata& 1.05\nl
11& Mg$_1$ & \nodata& \nodata& 0.92\nl
12& Mg$_2$ & \nodata& \nodata& 0.92\nl
13& Mg $b$ & 0.94& \nodata& 0.94\nl
14& Fe5270 & 0.75& \nodata& 0.75\nl
15& Fe5335 & 0.95& \nodata& 0.95\nl
16& Fe5406 & 0.88& \nodata& 0.88\nl
17& Fe5709 & \nodata& 0.94& 0.94\nl
18& Fe5782 & \nodata& 0.81& 0.81\nl
19& Na D   & \nodata& \nodata& 0.92\nl
20& TiO$_1$& \nodata& \nodata& 0.92\nl
21& TiO$_2$& \nodata& \nodata& 0.92\nl
\enddata
\tablenotetext{a}{These corrections adjust the assumed standard star
errors in Table 2 to produce the correct mean error level relative to
Gonz{\'a}lez (1993; G93) and the right balance among index errors internal
to the IDS data as described in Section 3.}
\end{deluxetable}

\clearpage

\begin{deluxetable}{lrrc}
\scriptsize
\tablecaption{Velocity dispersions used to correct the raw
indices\label{tbl4}}
\tablewidth{0pt}
\tablehead{\colhead{Name}&\colhead{$\sigma$}&\colhead{$\epsilon_{\sigma_v}$}&\colhead{Source}}
\startdata
A 569A&226&14&3\nl
IC 171&179&14&3\nl
IC 179&214&14&3\nl
IC 310&232&14&3\nl
IC 783&100&50&8\nl
IC 1131&104&20&5\nl
IC 1696&169&14&3\nl
IC 1907&238&14&3\nl
IC 2955&188&14&3\nl
IC 3303&100&50&8\nl
IC 3470&120&23&4\nl
IC 3652&100&50&8\nl
IC 3653&240&14&3\nl
IC 3672&100&50&8\nl
IC 4051&223&14&3\nl
NGC 80&296&14&3\nl
NGC 83&254&14&3\nl
NGC 128&198&12&5\nl
NGC 185&23&22&6\nl
NGC 194&208&14&3\nl
NGC 205&14&7&7\nl
NGC 221&77&3&1\nl
NGC 224&183&1&1\nl
NGC 227&268&14&3\nl
NGC 315&310&1&1\nl
NGC 379&245&14&3\nl
NGC 380&277&14&3\nl
NGC 382&153&14&3\nl
NGC 383&265&14&3\nl
NGC 385&180&14&3\nl
NGC 386&61&14&3\nl
NGC 392&261&14&3\nl
NGC 404&55&14&3\nl
NGC 410&321&14&3\nl
\tablebreak
NGC 474&171&13&4\nl
NGC 499&237&14&3\nl
NGC 501&163&14&3\nl
NGC 507&275&2&1\nl
NGC 524&275&10&2\nl
NGC 529&216&14&3\nl
\enddata
\tablecomments{Columns:\\
(1) Galaxy name.  See note, Table 6.\\
(2) Velocity dispersion, $\sigma$, in units of \kms.\\
(3) Fractional uncertainty of velocity dispersion, in percent.  Taken
from estimates in individual sources except source 7, whose
uncertainties were estimated to be 10\%, and this paper (source 8),
in which velocity dispersions and uncertainties are based on eye
estimates on comparison to galaxies with similar Mg$_2$ using the
Mg$_2$--$\sigma$ relation.\\
(4) Sources of velocity dispersion.  1=Gonz{\'a}lez (1993); 2=Faber et 
al.\ (1997); 3=Faber et al.\ (1989); 4=Whitmore, McElroy \& Tonry (1985); 
5=Dalle Ore et al.\ (1991); 6=Bender, Paquet \& Nieto (1991); 7=Peterson 
\& Caldwell (1993); 8=this paper (rough eye estimates; see text).}
\end{deluxetable}

\clearpage

\begin{deluxetable}{llrrrrr}
\tablecaption{Velocity dispersion correction polynomial coefficients\label{tbl5}}
\tablewidth{0pt}
\tablehead{
\colhead{$j$} & \colhead{Name} & \colhead{$c_0$} & \colhead{$c_1$} &
\colhead{$c_2$} & \colhead{$c_3$}}
\startdata
01 &  CN$_1$ &   1.000e+00&  3.333e$-$05&  2.222e$-$07& $-$7.105e$-$15\nl
02 &  CN$_2$ &   1.000e$+$00&  5.333e$-$05&  5.333e$-$07& $-$2.963e$-$10\nl
03 &  Ca4227 &   1.000e$+$00&  1.378e$-$04&  1.356e$-$06&  9.432e$-$09\nl
04 &  G4300  &   1.000e$+$00&  7.222e$-$05&  4.000e$-$07&  3.457e$-$10\nl
05 &  Fe4383 &   1.000e$+$00&  5.553e$-$06&  1.933e$-$06& $-$9.877e$-$10\nl
06 &  Ca4455 &   1.000e$+$00&  1.489e$-$04&  2.467e$-$06&  4.198e$-$09\nl
07 &  Fe4531 &   1.000e$+$00&  3.889e$-$05&  2.578e$-$06& $-$1.136e$-$09\nl
08 &  C$_2$4668 &   1.000e$+$00& $-$6.667e$-$06& 1.244e$-$06& $-$2.963e$-$10\nl
09 & H$\beta$&   1.000e$+$00&  7.444e$-$05&  2.667e$-$07&  1.136e$-$09\nl
10 &  Fe5015 &   1.000e$+$00&  9.667e$-$05&  2.578e$-$06& $-$1.926e$-$09\nl
11 &  Mg$_1$ &   1.000e$+$00& $-$2.223e$-$06& 5.333e$-$07& $-$4.938e$-$10\nl
12 &  Mg$_2$ &   1.000e$+$00&  3.444e$-$05& $-$4.445e$-$08&  2.469e$-$10\nl
13 &  Mg $b$ &   1.000e$+$00& $-$9.333e$-$05& 2.800e$-$06& $-$1.481e$-$09\nl
14 & Fe5270 &   1.000e$+$00&  4.000e$-$05&  2.667e$-$06& $-$1.481e$-$09\nl
15 &  Fe5335 &   1.000e$+$00& $-$5.667e$-$05&  5.444e$-$06&  2.963e$-$10\nl
16 &  Fe5406 &   1.000e$+$00& $-$6.778e$-$05&  4.956e$-$06&  7.901e$-$10\nl
17 &  Fe5709 &   1.000e$+$00&  2.111e$-$04&  6.222e$-$07&  4.839e$-$09\nl
18 &  Fe5782 &   1.000e$+$00&  1.033e$-$04&  2.867e$-$06&  5.926e$-$09\nl
19 &  Na D   &   1.000e$+$00&  5.222e$-$05&  2.000e$-$07&  1.975e$-$09\nl
20 & TiO$_1$ &   1.000e$+$00& $-$3.922e$-$04& 3.178e$-$06& $-$3.753e$-$09\nl
21 & TiO$_2$ &   1.000e$+$00& $-$8.889e$-$05& 7.111e$-$07& $-$7.901e$-$10\nl
\enddata
\end{deluxetable}

\clearpage

\begin{deluxetable}{llllll}
\tablecaption{Mean measurement differences\label{tblb1}}
\tablewidth{0pt}
\tablehead{
 & & \multicolumn{2}{c}{Eye$-$AUTOINDEX} &
\multicolumn{2}{c}{Published$-$AUTOINDEX} \\
 & & \multicolumn{2}{c}{(raw)} & \multicolumn{2}{c}{(K giant
standards)} \\
\colhead{$j$} & \colhead{Index} & \colhead{Galaxies\tablenotemark{a}}
& \colhead{All stars\tablenotemark{a}} & \colhead{Paper
V\tablenotemark{b}} & \colhead{This paper\tablenotemark{c}} }
\tablecolumns{6}
\startdata
01&CN$_1$&$-$0.002&$-$0.005&$-$0.007&$-$0.010\nl
04&G4300&$-$0.02&\phs0.01&$-$0.29&$-$0.21\nl
09&H$\beta$&$-$0.01&$-$0.02&$-$0.05&$-$0.03\nl
11&Mg$_1$&$-$0.001&$-$0.001&$-$0.007&$-$0.007\nl
12&Mg$_2$&\phs0.003&\phs0.000&$-$0.001&$-$0.002\nl
13&Mg $b$&\phs0.00&$-$0.04&$-$0.05&$-$0.01\nl
14&Fe5270&\phs0.00&$-$0.04&$-$0.04&$-$0.02\nl
15&Fe5335&$-$0.04&$-$0.04&$-$0.10&$-$0.10\nl
19&Na D&$-$0.04&\phs0.00&\phs0.03&\phs0.06\nl
20&TiO$_1$&$-$0.001&$-$0.001&\phs0.001&\phs0.001\nl
21&TiO$_2$&$-$0.001&$-$0.001&\phs0.000&\phs0.000\nl
\enddata
\tablenotetext{a}{Eye$-$AUTOINDEX raw values; run corrections and
velocity dispersion corrections have not been applied.}
\tablenotetext{b}{Correction onto Lick/IDS system as
determined for Paper V and applied to AUTOINDEX measurements of stars
there.  Based on nine K giant standard stars.}
\tablenotetext{c}{Repeat analysis of Paper V corrections based on same
nine K giant standard stars.  Values applied to all galaxies and
globular clusters in this paper.}
\end{deluxetable}

\begin{table}
\dummytable\label{tbl6a}
\end{table}

\begin{table}
\dummytable\label{tbl6b}
\end{table}

\begin{table}
\dummytable\label{tbl6c}
\end{table}

\begin{table}
\dummytable\label{tbl6d}
\end{table}

\clearpage

\begin{deluxetable}{llrrr}
\tablecaption{Multiplicative corrections to bring wide slit
observations onto $1\farcs4$ system\label{tbl7}}
\tablewidth{0pt}
\tablehead{
& & \colhead{3\farcs4} & \colhead{5\farcs4} & \colhead{7\farcs4}\\
\colhead{$j$} & \colhead{Name} & \colhead{correction} &
\colhead{correction} & \colhead{correction}}
\tablecolumns{5}
\startdata
 01 & CN$_1$ & 1.05 & 1.06 & 1.08\nl
 02 & CN$_2$ & 1.07 & 1.10 & 1.13\nl
 03 & Ca4227 & 1.00 & 1.00 & \nodata\nl
 04 & G4300  & 1.07 & 1.09 & 1.12\nl
 05 & Fe4383 & 1.06 & 1.13 & 1.20\nl
 06 & Ca4455 & 1.08 & 1.08 & \nodata\nl
 07 & Fe4531 & 1.07 & 1.07 & \nodata\nl
 08 & C$_2$4668& 1.03 & 1.10 & 1.17\nl
 09 & H$\beta$& 1.06 & 1.06 & \nodata\nl
 10 & Fe5015 & 1.07 & 1.07 & \nodata\nl
 11 & Mg$_1$ & 1.01 & 1.02 & 1.03\nl
 12 & Mg$_2$ & 1.00 & 1.01 & 1.02\nl
 13 & Mg $b$ & 1.04 & 1.11 & 1.18\nl
 14 & Fe5270 & 1.04 & 1.13 & 1.22\nl
 15 & Fe5335 & 1.03 & 1.03 & \nodata\nl
 16 & Fe5406 & 1.04 & 1.04 & \nodata\nl
 17 & Fe5709 & 1.06 & 1.06 & \nodata\nl
 18 & Fe5782 & 1.14 & 1.14 & \nodata\nl
 19 & Na D   & 0.99 & 1.01 & 1.02\nl
 20 & TiO$_1$& 1.00 & 1.07 & 1.15\nl
 21 & TiO$_2$& 0.95 & 0.99 & 1.03\nl
\enddata
\end{deluxetable}

\clearpage

\begin{deluxetable}{lll}
\tablecaption{Indices most affected by strong night-sky
emission\label{tbl8}}
\tablewidth{0pt}
\tablehead{
\colhead{Name} & \colhead{Contaminants} & \colhead{Location}}
\startdata
G4300&Hg I $\lambda$4358&sideband, for some redshifts\nl
Fe5406&Hg I $\lambda$5461&sideband, for some redshifts\nl
Fe5709&Na I $\lambda\lambda$5683,5688&sideband\nl
Fe5782&Hg I $\lambda$5770&sideband\nl
&Hg I $\lambda$5791&central bandpass\nl
\enddata
\end{deluxetable}

\end{document}


\begin{deluxetable}{lcrrrrrrrrrrrrrrrrrrrrr}
\tablenum{9}
\tablecaption{Fully corrected indices for off-nuclear locations
in galaxies observed through the standard aperture}
\tablewidth{0pt}
\tablehead{\colhead{Name}&&\colhead{CN$_1$}&\colhead{CN$_2$}&\colhead{Ca42}&\colhead{G}&\colhead{Fe43}&\colhead{Ca44}&\colhead{Fe45}&\colhead{C$_2$46}&\colhead{H$\beta$}&\colhead{Fe50}&\colhead{Mg$_1$}&\colhead{Mg$_2$}&\colhead{Mg $b$}&\colhead{Fe52}&\colhead{Fe53}&\colhead{Fe54}&\colhead{Fe57}&\colhead{Fe57}&\colhead{Na D}&\colhead{TiO$_1$}&\colhead{TiO$_2$}\\
\colhead{G}&\colhead{N}&\colhead{$\sigma$}&\colhead{$\sigma$}&\colhead{$\sigma$}&\colhead{$\sigma$}&\colhead{$\sigma$}&\colhead{$\sigma$}&\colhead{$\sigma$}&\colhead{$\sigma$}&\colhead{$\sigma$}&\colhead{$\sigma$}&\colhead{$\sigma$}&\colhead{$\sigma$}&\colhead{$\sigma$}&\colhead{$\sigma$}&\colhead{$\sigma$}&\colhead{$\sigma$}&\colhead{$\sigma$}&\colhead{$\sigma$}&\colhead{$\sigma$}&\colhead{$\sigma$}&\colhead{$\sigma$}}
\tablecolumns{23}
\startdata
\multicolumn{2}{l}{NGC221 5ew}&-0.001& 0.001& 1.83& 4.47& 5.33& 1.63& 3.81& 5.32& 1.84& 5.39& 0.076& 0.196& 2.90& 3.08& 2.21& 1.43& 0.91& 0.80& 2.77& 0.025& 0.052\nl
\quad 1066& 1& 0.004& 0.046& 0.60& 0.76& 1.33& 0.50& 0.87& 1.36& 0.46& 1.12& 0.013& 0.017& 0.49& 0.47& 0.54& 0.41& 0.39& 0.41& 0.52& 0.015& 0.013\nl
\multicolumn{2}{l}{NGC221 10ew}& 0.037& 0.054& 1.84& 4.81& 5.40& 1.59& 2.75& 5.67& 2.33& 5.31& 0.092& 0.207& 3.16& 2.67& 2.40& 1.44& 0.98& 1.10& 2.95& 0.037& 0.055\nl
\quad 1319& 1& 0.028& 0.039& 0.50& 0.64& 1.11& 0.42& 0.73& 1.14& 0.39& 0.94& 0.011& 0.014& 0.41& 0.39& 0.45& 0.34& 0.33& 0.35& 0.44& 0.012& 0.011\nl
\multicolumn{2}{l}{NGC221 15ew}& 0.003& 0.054& 1.26& 5.11&\nodata& 1.53& 0.68& 6.33& 2.04& 4.94& 0.088& 0.222& 3.17& 1.99& 1.27& 2.09& 1.32&\nodata& 3.04& 0.051& 0.067\nl
\quad  405& 1& 0.024& 0.114& 1.48& 1.89&\nodata& 1.25& 2.16& 3.36& 1.14& 2.77& 0.034& 0.042& 1.22& 1.16& 1.29& 1.01& 0.97&\nodata& 1.29& 0.036& 0.032\nl
\multicolumn{2}{l}{NGC224 5ew}& 0.143& 0.161& 1.77& 5.55& 5.79& 1.22& 3.37& 7.58& 1.92& 5.98& 0.147& 0.309& 5.09& 2.70& 2.11& 1.75& 1.07& 0.72& 4.77& 0.041& 0.092\nl
\quad 1575& 1& 0.029& 0.034& 0.48& 0.57& 1.02& 0.41& 0.68& 1.02& 0.34& 0.89& 0.010& 0.012& 0.40& 0.37& 0.46& 0.34& 0.31& 0.34& 0.39& 0.011& 0.009\nl
\multicolumn{2}{l}{NGC224 10ew}& 0.126& 0.139& 1.85& 5.17& 3.55& 1.47& 3.30& 6.58& 1.95& 5.21& 0.145& 0.292& 4.41& 3.08& 2.44& 1.96& 1.04& 0.91& 4.98& 0.038& 0.084\nl
\quad 1986& 1& 0.024& 0.029& 0.40& 0.48& 0.85& 0.34& 0.57& 0.86& 0.29& 0.74& 0.009& 0.010& 0.33& 0.32& 0.39& 0.29& 0.26& 0.28& 0.33& 0.009& 0.008\nl
\multicolumn{2}{l}{NGC224 15ew}& 0.104& 0.136& 1.42& 4.64& 5.61& 0.97& 2.44& 7.40& 1.56& 4.39& 0.144& 0.302& 4.27& 2.79& 2.14& 2.12& 1.37& 0.90& 4.60& 0.025& 0.081\nl
\quad 1270& 1& 0.033& 0.040& 0.56& 0.67& 1.19& 0.47& 0.79& 1.20& 0.40& 1.02& 0.012& 0.015& 0.45& 0.43& 0.52& 0.40& 0.36& 0.39& 0.45& 0.013& 0.011\nl
\multicolumn{2}{l}{NGC224 20ew}& 0.108& 0.207& 2.20& 4.94& 4.34& 1.90& 3.90& 6.64& 1.19& 6.55& 0.139& 0.308& 4.54& 2.93& 1.61& 1.49& 0.61& 1.21& 5.38& 0.049& 0.104\nl
\quad  899& 1& 0.045& 0.054& 0.75& 0.90& 1.60& 0.63& 1.07& 1.62& 0.54& 1.38& 0.017& 0.020& 0.61& 0.58& 0.68& 0.52& 0.48& 0.52& 0.61& 0.017& 0.015\nl
\multicolumn{2}{l}{NGC224 35ew}& 0.077& 0.115& 1.33& 5.16& 5.00& 1.77& 2.86& 6.87& 1.25& 4.63& 0.127& 0.267& 4.17& 2.33& 2.44& 1.58& 0.76& 0.87& 4.17& 0.039& 0.078\nl
\quad 2689& 4& 0.015& 0.019& 0.27& 0.32& 0.57& 0.23& 0.38& 0.58& 0.19& 0.50& 0.006& 0.007& 0.23& 0.21& 0.27& 0.19& 0.17& 0.19& 0.22& 0.006& 0.005\nl
\multicolumn{2}{l}{NGC3115 5mj}&\nodata&\nodata& 1.07& 4.81& 4.53& 1.51& 3.39& 8.30& 2.27& 4.24& 0.164& 0.310& 4.34& 3.08& 2.83& 2.34& 0.86& 0.90& 4.85& 0.054& 0.096\nl
\quad 1171& 1&\nodata&\nodata& 0.71& 0.74& 1.37& 0.60& 0.94& 1.37& 0.46& 1.21& 0.014& 0.016& 0.56& 0.53& 0.74& 0.56& 0.43& 0.51& 0.53& 0.015& 0.012\nl
\multicolumn{2}{l}{NGC3115 7mj}& 0.079& 0.118& 0.45& 5.95& 7.70& 1.95& 4.16& 8.45& 2.01& 6.20& 0.150& 0.298& 4.76& 4.01& 3.46& 1.46& 0.82& 0.58& 4.32& 0.069& 0.070\nl
\quad 2222& 1& 0.022& 0.028& 0.43& 0.48& 0.91& 0.41& 0.62& 0.88& 0.29& 0.82& 0.009& 0.010& 0.40& 0.38& 0.57& 0.34& 0.27& 0.31& 0.34& 0.009& 0.008\nl
\multicolumn{2}{l}{NGC3115 10mj}& 0.096& 0.146& 0.72& 5.03& 4.22& 1.53& 3.12& 7.59& 1.65& 5.51& 0.146& 0.280& 4.04& 3.05& 2.58& 1.35& 0.82& 0.98& 4.15& 0.040& 0.078\nl
\quad 2131& 2& 0.021& 0.026& 0.40& 0.44& 0.80& 0.36& 0.55& 0.81& 0.26& 0.75& 0.008& 0.009& 0.35& 0.33& 0.47& 0.31& 0.25& 0.30& 0.31& 0.009& 0.007\nl
\multicolumn{2}{l}{NGC3115 15mj}&\nodata&\nodata& 1.75& 4.87& 7.81& 1.08& 4.68& 9.46& 2.02& 3.90& 0.169& 0.276& 4.06& 2.27& 1.32& 1.45& 0.97& 1.89& 4.30& 0.059& 0.085\nl
\quad  643& 1&\nodata&\nodata& 1.18& 1.26& 2.32& 0.97& 1.58& 2.31& 0.77& 2.01& 0.023& 0.027& 0.89& 0.85& 1.07& 0.83& 0.71& 0.85& 0.87& 0.025& 0.021\nl
\multicolumn{2}{l}{NGC3115 28mj}& 0.067& 0.116&-0.16& 6.01& 2.80& 1.58& 2.73& 8.10& 1.82& 6.42& 0.118& 0.247& 4.26& 3.67& 3.75& 0.78&-0.19& 0.72& 2.40& 0.020& 0.095\nl
\quad  767& 1& 0.050& 0.064& 0.97& 1.08& 1.94& 0.82& 1.32& 1.95& 0.65& 1.72& 0.019& 0.023& 0.76& 0.73& 1.00& 0.69& 0.59& 0.69& 0.74& 0.021& 0.018\nl
\multicolumn{2}{l}{NGC3379 10ew}&\nodata&\nodata& 0.45& 4.52& 5.11& 2.19& 3.12& 6.66& 1.44& 3.55& 0.156& 0.295& 4.07& 2.61& 2.04& 0.93& 0.79& 1.29& 3.42& 0.028& 0.065\nl
\quad 1011& 1&\nodata&\nodata& 0.70& 0.81& 1.47& 0.60& 0.99& 1.48& 0.49& 1.27& 0.015& 0.018& 0.57& 0.54& 0.67& 0.49& 0.44& 0.50& 0.56& 0.016& 0.014\nl
\multicolumn{2}{l}{NGC3379 20ns}&\nodata&\nodata& 0.78& 5.68& 6.67& 2.77& 3.11& 5.52& 3.15& 6.73& 0.130& 0.279& 4.53& 2.74& 3.46& 1.46& 1.64&-0.86& 4.38& 0.014& 0.026\nl
\quad  369& 1&\nodata&\nodata& 1.79& 2.12& 3.78& 1.52& 2.54& 3.81& 1.29& 3.26& 0.038& 0.046& 1.43& 1.37& 1.71& 1.26& 1.14& 1.26& 1.44& 0.043& 0.035\nl
\multicolumn{2}{l}{NGC4111 8mj}& 0.011& 0.056& 0.80& 5.18& 4.89& 0.88& 2.94& 7.36& 2.19& 4.62& 0.086& 0.207& 3.13& 2.69& 3.33& 1.58& 1.04& 0.77& 4.96& 0.038& 0.075\nl
\quad 2248& 1& 0.013& 0.026& 0.36& 0.44& 0.78& 0.30& 0.52& 0.79& 0.27& 0.67& 0.008& 0.010& 0.29& 0.28& 0.35& 0.25& 0.23& 0.25& 0.30& 0.009& 0.007\nl
\multicolumn{2}{l}{NGC4472 10ew}&\nodata&\nodata& 1.04& 4.81& 4.51& 1.72& 3.00& 5.20& 1.77& 6.86& 0.148& 0.316& 4.81& 2.00& 2.78& 1.30& 0.67& 0.53& 4.22& 0.053& 0.061\nl
\quad 1045& 1&\nodata&\nodata& 0.78& 0.82& 1.50& 0.66& 1.03& 1.49& 0.50& 1.36& 0.015& 0.017& 0.62& 0.56& 0.80& 0.56& 0.47& 0.55& 0.57& 0.016& 0.013\nl
\multicolumn{2}{l}{NGC4472 20ns}& 0.084& 0.122& 0.35& 4.96& 5.86& 1.47& 3.08& 6.65& 0.73& 4.29& 0.161& 0.306& 4.41& 1.57& 1.46& 1.57& 0.83& 0.28& 3.54& 0.035& 0.101\nl
\quad 1022& 2& 0.038& 0.048& 0.74& 0.80& 1.48& 0.64& 1.01& 1.47& 0.47& 1.29& 0.015& 0.017& 0.60& 0.54& 0.70& 0.55& 0.45& 0.52& 0.56& 0.016& 0.013\nl
\enddata
\tablecomments{
This table gives off-nuclear observations of galaxies through the
standard slitwidth (1\farcs4 by 4\farcs0).  The name consists of the
NGC number followed by an offset or a position.  The offsets are
indicated by a number (the offset in arcseconds from the nucleus)
followed by a two-letter code: ``ew'' means that the offset is along
the east-west line; ``ns'' is along the north-south line; ``mj'' is
along the major axis; ``mn'' is along the minor axis.  Signs are not
given because the data are sometimes averaged about the nucleus.  A
position code $x,y$ denotes an offset along the major axis by the
amount $x$ and an offset along the minor axis by the amount $y$.
These are likewise not signed for the same reason.}
\end{deluxetable}


\begin{deluxetable}{lccrrrrrrrrrrrrrrrrrrrrr}
\tablenum{10}
\tablecaption{Fully corrected indices for non-standard aperture observations
of galaxies and M31 globular clusters}
\tablewidth{0pt}
\tablehead{\colhead{Name}&&\colhead{width}&\colhead{CN$_1$}&\colhead{CN$_2$}&\colhead{Ca42}&\colhead{G}&\colhead{Fe43}&\colhead{Ca44}&\colhead{Fe45}&\colhead{C$_2$46}&\colhead{H$\beta$}&\colhead{Fe50}&\colhead{Mg$_1$}&\colhead{Mg$_2$}&\colhead{Mg $b$}&\colhead{Fe52}&\colhead{Fe53}&\colhead{Fe54}&\colhead{Fe57}&\colhead{Fe57}&\colhead{Na D}&\colhead{TiO$_1$}&\colhead{TiO$_2$}\\
\colhead{G}&\colhead{N}&\colhead{length}&\colhead{$\sigma$}&\colhead{$\sigma$}&\colhead{$\sigma$}&\colhead{$\sigma$}&\colhead{$\sigma$}&\colhead{$\sigma$}&\colhead{$\sigma$}&\colhead{$\sigma$}&\colhead{$\sigma$}&\colhead{$\sigma$}&\colhead{$\sigma$}&\colhead{$\sigma$}&\colhead{$\sigma$}&\colhead{$\sigma$}&\colhead{$\sigma$}&\colhead{$\sigma$}&\colhead{$\sigma$}&\colhead{$\sigma$}&\colhead{$\sigma$}&\colhead{$\sigma$}&\colhead{$\sigma$}}
\tablecolumns{24}
\startdata
M31 V29&&3\farcs4& 0.112& 0.156& 0.41& 7.87& 7.29& 1.94& 2.28& 4.39& 1.78& 4.38& 0.081& 0.212& 4.54& 1.97& 0.97& 1.00& 0.74&\nodata& 3.26& 0.031& 0.068\nl
\quad 1630& 1&4\arcsec& 0.027& 0.033& 0.42& 0.54& 0.93& 0.35& 0.61& 0.96& 0.32& 0.78& 0.010& 0.012& 0.35& 0.33& 0.35& 0.28& 0.27&\nodata& 0.37& 0.011& 0.009\nl
M31 V92&&2\farcs2&-0.025& 0.015& 1.08& 4.39& 3.00& 0.69& 3.55& 1.33& 1.79& 2.28& 0.026& 0.104& 1.76& 1.68& 0.99& 0.31& 0.87&\nodata& 1.60& 0.020& 0.011\nl
\quad 2314& 1&4\arcsec& 0.039& 0.026& 0.33& 0.42& 0.74& 0.28& 0.48& 0.76& 0.26& 0.62& 0.006& 0.009& 0.28& 0.26& 0.28& 0.23& 0.22&\nodata& 0.30& 0.009& 0.007\nl
IC783&&3\farcs4&-0.006& 0.018& 1.14& 2.48& 5.31& 1.92& 3.58& 3.94& 2.75& 4.26& 0.040& 0.120& 1.40& 1.89& 1.52& 0.74& 0.94& 1.04& 1.69& 0.018&\nodata\nl
\quad 1018& 1&4\arcsec& 0.060& 0.048& 0.63& 0.76& 1.39& 0.54& 0.93& 1.42& 0.48& 1.19& 0.013& 0.017& 0.52& 0.50& 0.57& 0.44& 0.41& 0.44& 0.55& 0.016&\nodata\nl
IC1131&&3\farcs4& 0.024& 0.115& 1.86&\nodata& 3.62&-0.03& 3.13& 3.76& 3.07& 2.76& 0.079& 0.211& 2.02& 3.61& 1.89&\nodata&\nodata& 0.54& 2.86& 0.025& 0.050\nl
\quad  735& 1&6\arcsec& 0.042& 0.065& 0.85&\nodata& 1.88& 0.72& 1.24& 1.91& 0.65& 1.59& 0.019& 0.023& 0.70& 0.67& 0.77&\nodata&\nodata& 0.59& 0.73& 0.021& 0.018\nl
IC3303&&5\farcs4&-0.120&-0.141&-0.81& 0.02& 1.83& 0.27& 0.78& 1.65& 3.06& 1.01& 0.021& 0.080& 0.96& 1.17&-0.20& 0.57& 0.09&\nodata& 0.48& 0.030& 0.033\nl
\quad 1183& 1&6\arcsec& 0.042& 0.042& 0.56& 0.07& 1.23& 0.47& 0.81& 1.25& 0.43& 1.04& 0.010& 0.015& 0.45& 0.43& 1.07& 0.38& 0.36&\nodata& 0.54& 0.014& 0.012\nl
IC3470&&3\farcs4& 0.107& 0.146& 1.14& 3.83& 2.47& 0.22& 4.42& 1.33& 2.13& 7.12& 0.051& 0.179& 2.47& 2.26& 1.50& 0.63& 1.54& 0.96& 1.29& 0.030& 0.047\nl
\quad  986& 1&4\arcsec& 0.041& 0.050& 0.66& 0.81& 1.44& 0.56& 0.96& 1.47& 0.50& 1.24& 0.014& 0.018& 0.54& 0.52& 0.59& 0.46& 0.43& 0.46& 0.58& 0.016& 0.014\nl
IC3652&&5\farcs4&-0.149&-0.140& 1.14& 4.12& 1.59& 1.97& 4.21& 2.59& 2.48& 2.75& 0.071& 0.137& 2.17& 4.61& 3.96& 1.41& 1.19&\nodata& 0.69& 0.019& 0.049\nl
\quad  838& 1&6\arcsec& 0.056& 0.057& 0.76& 0.94& 1.66& 0.64& 1.10& 1.70& 0.58& 1.41& 0.017& 0.021& 0.62& 0.61& 0.73& 0.52& 0.50&\nodata& 0.70& 0.019& 0.016\nl
IC3653&&3\farcs4& 0.028& 0.070& 1.61& 5.57& 5.94& 1.62& 5.41& 7.00& 2.11& 6.61& 0.091& 0.251& 4.48& 3.73& 1.87& 2.82& 1.76& 1.08& 2.91& 0.036& 0.051\nl
\quad 1393& 1&4\arcsec& 0.025& 0.038& 0.60& 0.64& 1.17& 0.50& 0.81& 1.16& 0.39& 1.04& 0.011& 0.013& 0.46& 0.45& 0.56& 0.46& 0.37& 0.42& 0.44& 0.013& 0.011\nl
IC3672&&5\farcs4& 0.017& 0.038& 1.33& 3.85& 2.08& 2.29& 2.82& 7.05& 2.27& 5.59& 0.055& 0.210& 3.92& 2.39& 2.89& 1.67& 0.26&\nodata& 2.25& 0.044& 0.056\nl
\quad  898& 1&6\arcsec& 0.031& 0.054& 0.71& 0.88& 1.56& 0.60& 1.03& 1.59& 0.54& 1.33& 0.015& 0.019& 0.59& 0.56& 0.67& 0.49& 0.46&\nodata& 0.61& 0.017& 0.015\nl
\multicolumn{2}{l}{NGC128 35mj}&3\farcs4& 0.163& 0.179&-0.19&\nodata& 6.15& 0.46& 2.72& 4.33& 2.21& 7.20& 0.116& 0.265& 4.67& 1.96& 3.14&\nodata&\nodata& 1.45& 5.08& 0.046&\nodata\nl
\quad  835& 1&16\arcsec& 0.050& 0.058& 0.80&\nodata& 1.72& 0.68& 1.15& 1.73& 0.59& 1.49& 0.018& 0.021& 0.65& 0.62& 0.77&\nodata&\nodata& 0.57& 0.66& 0.019&\nodata\nl
NGC185&&3\farcs4&-0.138&-0.102&-0.28& 1.60&\nodata& 0.04& 2.46& 1.44& 2.40& 2.76& 0.032& 0.106& 1.34& 1.40& 1.23& 0.80& 1.00&\nodata& 2.27& 0.014& 0.008\nl
\quad 1415& 1&16\arcsec& 0.036& 0.036& 0.47& 0.55&\nodata& 0.39& 0.68& 1.07& 0.36& 0.87& 0.010& 0.013& 0.39& 0.36& 0.40& 0.32& 0.31&\nodata& 0.41& 0.012& 0.010\nl
\multicolumn{2}{l}{NGC185 10ew}&3\farcs4&-0.132&-0.135& 0.53& 2.25&\nodata& 1.76& 2.39& 3.17& 1.49& 3.41& 0.019& 0.125& 2.72&-0.14& 1.10& 0.73& 1.23&\nodata& 3.11& 0.014& 0.059\nl
\quad  422& 1&16\arcsec& 0.108& 0.109& 1.40& 1.72&\nodata& 1.18& 2.05& 3.22& 1.09& 2.62& 0.026& 0.040& 1.17& 1.31& 1.20& 0.95& 0.92&\nodata& 1.23& 0.037& 0.031\nl
\multicolumn{2}{l}{NGC185 25ew}&3\farcs4&-0.170&-0.074& 0.27& 4.08&\nodata&-0.75& 4.56& 1.59& 2.32& 4.77& 0.028& 0.096& 2.06& 2.29& 1.34& 0.27& 1.37&\nodata& 1.29& 0.022& 0.069\nl
\quad  557& 1&16\arcsec& 0.081& 0.083& 1.07& 1.36&\nodata& 0.90& 1.57& 2.46& 0.84& 2.00& 0.021& 0.030& 0.89& 0.84& 0.93& 0.73& 0.70&\nodata& 0.97& 0.028& 0.024\nl
\multicolumn{2}{l}{NGC185 35ew}&3\farcs4&-0.126&-0.097& 0.42& 3.22&\nodata& 1.54& 2.61& 4.78& 3.38& 1.02& 0.057& 0.096& 1.45& 1.01& 2.33&-0.61& 0.50&\nodata& 0.62& 0.011&-0.005\nl
\quad  492& 1&16\arcsec& 0.093& 0.094& 1.21& 1.52&\nodata& 1.01& 1.77& 2.77& 0.95& 2.26& 0.027& 0.034& 1.00& 0.94& 1.08& 0.82& 0.79&\nodata& 1.15& 0.033& 0.026\nl
\multicolumn{2}{l}{NGC205 17mj}&3\farcs4&-0.155&-0.103& 0.26& 0.88& 3.45& 0.47& 1.34& 0.29& 3.12& 2.23& 0.017& 0.096& 1.43& 1.73& 1.06& 0.94& 0.94&\nodata& 1.22& 0.004& 0.022\nl
\quad 1719& 1&16\arcsec& 0.030& 0.031& 0.40& 0.43& 0.90& 0.34& 0.59& 0.92& 0.31& 0.75& 0.007& 0.011& 0.33& 0.32& 0.34& 0.27& 0.26&\nodata& 0.36& 0.013& 0.009\nl
\multicolumn{2}{l}{NGC205 54mj}&3\farcs4&-0.080&-0.059& 0.72& 2.27&\nodata& 0.76& 2.32& 1.60& 2.34&\nodata& 0.013& 0.090& 1.55& 2.17& 1.11& 0.86&-0.06&\nodata& 1.03& 0.001& 0.017\nl
\quad 1521& 1&16\arcsec& 0.036& 0.034& 0.44& 0.54&\nodata& 0.37& 0.64& 1.01& 0.34&\nodata& 0.007& 0.012& 0.37& 0.35& 0.38& 0.30& 0.29&\nodata& 0.40& 0.011& 0.010\nl
\multicolumn{2}{l}{NGC221 25ns}&1\farcs4&-0.025& 0.011& 0.95& 4.68&\nodata& 1.43& 3.31& 5.03& 1.99& 4.47& 0.070& 0.184& 3.42& 1.83& 1.47& 1.67& 1.01& 1.19& 2.82& 0.045& 0.058\nl
\quad 1998& 2&16\arcsec& 0.039& 0.026& 0.33& 0.42&\nodata& 0.28& 0.49& 0.76& 0.26& 0.62& 0.007& 0.009& 0.28& 0.26& 0.29& 0.23& 0.22& 0.23& 0.29& 0.008& 0.007\nl
\multicolumn{2}{l}{NGC221 45ns}&1\farcs4&-0.023& 0.011& 1.00& 4.31& 5.80& 1.24& 2.92& 6.50& 2.18& 3.99& 0.073& 0.178& 3.26& 2.64& 1.69& 1.51& 1.24& 1.07& 2.82& 0.067& 0.064\nl
\quad 1724& 2&16\arcsec& 0.046& 0.029& 0.38& 0.48& 0.84& 0.32& 0.55& 0.86& 0.29& 0.71& 0.009& 0.011& 0.31& 0.30& 0.33& 0.26& 0.25& 0.26& 0.33& 0.009& 0.008\nl
\enddata
\tablecomments{
This table contains all observations of galaxies (plus two M31
globular clusters) made through non-standard apertures (raster scans
of Milky Way globular clusters are given in Table 7).  Slitwidth and
length are given in the third column.  ``Scan'' denotes a raster scan
made with a $1\farcs4\times16\arcsec$ slit covering a square aperture
16\arcsec\ by 16\arcsec\ centered on the nucleus.  These scans
resemble the larger raster scans of globulars in Table 7 and have the
same spectral resolution as standard-slitwidth scans (1\farcs4).  They
were taken to determine the aperture correction to nuclear
$\sigma$ (see Davies et al.~1987).
\\
The naming convention for off-nuclear locations in this table is the
same as in Table 8.  Special names are:
\\
M31 V29 and M31 V92: Two M31 globular clusters observed with a
wide slit.  See notes to Table 7 for further information.
\\
NGC 3379 61se: Aperture located 61\arcsec\ south-east of the nucleus.}
\end{deluxetable}